\documentclass[12pt,english]{article} 
\usepackage[T1]{fontenc}
\usepackage{a4wide}
\usepackage{babel}
\usepackage{psfrag}
\usepackage[dvips]{graphicx}
\usepackage{amssymb}
\usepackage{fancyhdr}

\makeatletter
\newcommand{\lyxaddress}[1]{
  \par {\raggedright #1 
  \vspace{1.4em}
  \noindent\par}}
\newcommand{\ud}{\mathrm d}
\renewcommand\theequation{\hbox{\normalsize\arabic{section}.\arabic{equation}}}
\@addtoreset{equation}{section}
\newcommand{\nord}[1]{\ :\!#1\!:}
\makeatother
\begin{document}

\title{
\begin{flushright}
\normalsize{
{ITP--Budapest Report No. 557\\
KCL--MTH--00--21
}}
\end{flushright}
\vspace{1cm}
\textbf{\large The \protect\( k\protect \)-folded sine--Gordon model in finite
volume }\large }
\date{19th April 2000}
\author{Z.~Bajnok\protect\( ^{(a)}\protect \), L.~Palla\protect\( ^{(a)}\protect \)\thanks{
Corresponding author's e-mail: palla@ludens.elte.hu
}, G.~Tak\'{a}cs\protect\( ^{(b)}\protect \) and F.~W\'{a}gner\protect\( ^{(a)}\protect \)}

\maketitle

\lyxaddress{\centering \protect\( ^{(a)}\protect \)\emph{Institute for Theoretical Physics
}\\
\emph{E\"{o}tv\"{o}s University} \emph{}\\
\emph{H-1117 Budapest, P\'{a}zm\'{a}ny P.~s\'{e}t\'{a}ny 1/A, Hungary}
\\
\protect\( ^{(b)}\protect \)\emph{Department of Mathematics}\\
\emph{Kings College London,}  \emph{}\\
\emph{Strand, London WC2R 2LS, UK}}

\begin{abstract}
We consider the $k$-folded sine--Gordon model, obtained from the usual
version by identifying the scalar field after $k$ periods of the
cosine potential. We examine (1) the ground state energy split, (2)
the lowest lying multi-particle state spectrum and (3) vacuum
expectation values of local fields in finite spatial volume, combining
the Truncated Conformal Space Approach, the method of the Destri--de
Vega nonlinear integral equation (NLIE) and semiclassical instanton
calculations. We show that the predictions of all these different
methods are consistent with each other and in particular provide
further support for the NLIE method in the presence of a twist
parameter. It turns out that the model provides an optimal laboratory
for examining instanton contributions beyond the dilute instanton gas
approximation. We also provide evidence for the exact formula for the
vacuum expectation values conjectured by Lukyanov and Zamolodchikov.
\end{abstract}

{\par\centering PACS codes: 11.55.Ds, 11.30.Na, 11.10.Kk \\[.5cm]
}

\clearpage

\section{Introduction \label{sec:introduction}}

Completely integrable $2$ dimensional quantum field theories are
important sources of nonperturbative information but are rather
rare. Thus there is a considerable interest in modifications or
deformations that generate new integrable models from an old one. The
extensively studied examples of such modifications include the
integrable deformations of CFTs \cite{Zam1} and the class of boundary
integrable theories \cite{GZ} where the new boundary conditions
preserve integrability.

In this paper we consider a slightly different integrable modification
of sine--Gordon (SG) model, which was first mentioned in \cite{Sw} and
discussed in some detail in \cite{KM3}.  In this model the period of
the sine--Gordon field $\phi$ is $k$ times the period of the potential
for some generic $k\in {\mathbb {N}}$\footnote{This procedure can be
carried out in any scalar field theory---not necessarily
integrable---in which the scalar potential is a periodic function.}.
Thus the new model---which we denote by $\mathrm{SG}(\beta
,k)$---contains the folding number $k$ as a parameter in addition to
the usual coupling constant $\beta$. All properties of the ordinary SG
model ($\mathrm{SG}(\beta ,1)$) that rely only on the local properties
of the field $\phi $, like classical integrability, or the existence
of conserved higher spin quantities, will also hold for
$\mathrm{SG}(\beta ,k)$. Thus we have a rather clear intuitive picture
of the various (infinite volume) excitations of $\mathrm{SG}(\beta
,k)$, at least for $\beta^2<8\pi$.  This picture predicts the exact
$S$-matrices of the scattering among the particles corresponding to
these excitations in terms of the well known $S$-matrix of SG.
Nevertheless the spectrum of $\mathrm{SG}(\beta ,k)$ is not identical
to that of the ordinary SG, reflecting the importance of the boundary
conditions imposed. In particular the quantum theory
$\mathrm{SG}(\beta ,k)$ has a $k$-fold degenerate vacuum corresponding
to the \lq unidentified' minima of the potential. As a consequence
$\mathrm{SG}(\beta ,k)$ contains kinks, i.e.~particles that
interpolate between different vacua and have nontrivial restrictions
on their multi-particle Hilbert space. Since these restrictions are
$k$ dependent they do give rise to differences between the theories
with different $k$, that can manifest themselves in their finite
volume spectra.

In this paper we investigate $\mathrm{SG}(\beta ,k)$ in finite volume. 
We study three sets of problems in some detail, namely
the split in the vacuum energy levels, the spectrum of the low lying
multi-particle states, and finally the vacuum expectation values of
exponential fields. In all three cases we compare the theoretical
predictions with numerical data obtained by 
using the Truncated Conformal Space Approach (TCSA) \cite{yz_tcs}. 

We derive the split of the vacuum energy levels by two methods, on the
one hand we obtain it from the nonlinear integral equation (NLIE)
\cite{kbp}, \cite{ddv} appropriately generalized to describe
$\mathrm{SG}(\beta ,k)$, while on the other we perform an instanton
calculation. These two methods give identical results for the leading
part of the split, and the TCSA data show an excellent agreement with
this prediction. This agreement confirms the correctness of the
generalized NLIE.  We also show that the TCSA data make it possible to
extract and analyze the nonleading part of the split.

We determine the volume dependence of the energy levels of the low
lying multi-particle states by using the formalism of \cite{KM3}. This
method relies heavily on the conjectured $S$-matrices describing the
mutual scatterings in the multi-particle states.  For large volumes
this simpler and approximate method can be shown to give identical
results to the exact NLIE, and we chose it because it has a clear
interpretation in terms of the spectrum of the model. The---$k$
dependent---degeneracies and the volume dependence of the
multi-particle energy levels obtained this way agree very well with
the TCSA data, and this agreement confirms that the conjectured
$S$-matrices are indeed the correct ones.

The study of the vacuum expectation values of exponential fields is
motivated by the fact that in \cite{lz_vevs} an explicit formula was proposed
for this quantity at least in the ordinary sine--Gordon theory in
infinite volume. We translate this explicit expression into the finite
volume $\mathrm{SG}(\beta ,k)$, and compare it to the expectation values
measured in the ground states found by the TCSA.  The TCSA data are
good enough to distinguish between the semiclassical and exact
expressions given in \cite{lz_vevs}, favoring the latter.  The agreement we find
proves two things. On the one hand it confirms the expression given in
\cite{lz_vevs}, on the other it shows that the expectation, that everything,
which in sine--Gordon theory follows only from the local properties of
the scalar field and the Lagrangian, remains true also in the \(
k\)-folded model, is indeed correct.

The paper is organized as follows: in Section 2 we describe the
Lagrangian, the symmetries and the local operators of $\mathrm{SG}(\beta ,k)$, 
and recall the basic properties of TCSA as applied to this
model. Section 3 contains the generalization of the NLIE. We investigate
the vacuum structure of  $\mathrm{SG}(\beta ,k)$ in finite volume in
Section 4. Section 5 is devoted to the study of the multi-particle energy
levels, and we analyze the vacuum expectation values of exponential
fields in Section 6. We make our conclusions in Section 7. The paper is
closed by an appendix where we describe in detail the
computation of a determinant needed to complete the instanton
calculation of the split between  the vacuum energy levels.

\section{The \protect\( k\protect \)-folded sine--Gordon model \label{sec:the_model}}

\subsection{\protect\( \mathrm{SG}(\beta ,k)\protect \): Lagrangian and symmetries \label{subsec:lagrangian}}

The action of sine--Gordon theory in a finite spatial volume \( L \) is
\begin{equation}
\label{sG_action}
{\cal A}=\int ^{\infty }_{-\infty }\ud t\int ^{L/2}_{-L/2}\ud x\left( \frac{1}{2}\partial _{\mu }\varphi \partial ^{\mu }\varphi +\frac{\mu _{0}^{2}}{\beta ^{2}}\left( \cos \beta \varphi -1\right) \right) \,.
\end{equation}
For later convenience, we also define a new parameter \( p \) with
\[
p=\frac{\beta ^{2}}{8\pi -\beta ^{2}}\: .\]
To define the \( k \)-folded theory \( \mathrm{SG}(\beta ,k) \) we take the sine--Gordon
field \( \varphi  \) as an angular variable with the period 
\begin{figure}
\psfrag{f}{$\varphi$}
\psfrag{V}{$V(\varphi)$}
\psfrag{0}{$0$}
\psfrag{1}{$\frac{2\pi}{\beta}$}
\psfrag{2}{$\frac{4\pi}{\beta}$}
\psfrag{3}{$\frac{6\pi}{\beta}$}
\centering \includegraphics[width=5cm]{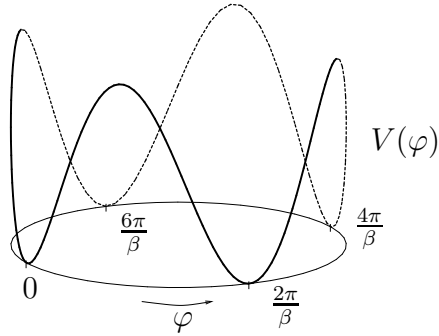}
\caption{The potential with the identification (\ref{period}) for $k=4$}
\label{kfoldpot}
\end{figure}

\begin{equation}
\label{period}
\varphi \, \sim \, \varphi +\frac{2\pi }{\beta }k\,.
\end{equation}
This implies the following quasi-periodic boundary condition for the field:
\begin{equation}
\label{quasi_periodic}
\varphi (x+L,t)=\varphi (x,t)+\frac{2\pi }{\beta }km\,,\quad m\in {\mathbb Z}\,.
\end{equation}
The classical ground states are easily obtained:
\begin{equation}
\label{classical_vacua}
\varphi_n =\frac{2\pi }{\beta }n\: ,\qquad n=0,\ldots ,k-1
\end{equation}
which shows that the condition (\ref{period}) corresponds to
identifying the minima of the cosine potential with a period \(k\)
(see Fig.~\ref{kfoldpot}).  All classical solutions of the ordinary SG
model are also solutions of \( \mathrm{SG}(\beta ,k) \), as the
equations of motion are identical. However the static soliton
solution, which, in the pure SG framework, on account of the
identification \( \varphi \equiv \varphi +2\pi/\beta \),
interpolates between the \emph{same} minimum, is now connecting
\emph{different}, \emph{neighbouring} minima; i.e.~the SG soliton
becomes a kink in \( \mathrm{SG}(\beta ,k) \).

In the infinite volume (\( L=\infty \)) quantum theory these
$\varphi_n$ correspond to the vacuum states \( \left| n\right\rangle
\) which have the property
\begin{equation}
\label{n_vacua}
\left\langle n\right| \varphi (x,t)\left| n\right\rangle =\frac{2\pi }{\beta }n\,.
\end{equation}
These states are all degenerate in the classical theory and also at quantum
level when \( L=\infty  \); however, tunnelling lifts the degeneracy in finite
volume \( L<\infty  \). 

Let us now examine the relevant symmetries of the action. One can
define a \( {\mathbb Z}_{k} \) group action generated by a unitary
operator \( T \) in the following way:
\begin{equation}
\label{T_symmetry}
T\varphi (x,t)T^{-1}=\varphi (x,t)-\frac{2\pi }{\beta }\,.
\end{equation}
Then we have
\begin{equation}\label{Tdef}
T\left| n\right\rangle =\left| n+1\,\bmod\, k\right\rangle
\end{equation} and the Hamiltonian
\begin{equation}
\label{sG_Hamiltonian}
H=\int ^{L/2}_{-L/2}\ud x\left( \frac{1}{2}\pi ^{2}+\frac{1}{2}\left( \partial _{x}\varphi \right) ^{2}+\frac{\mu_{0}^{2}}{\beta ^{2}}\left( 1-\cos \beta \varphi \right) \right) \, ,\qquad \pi (x,t)=\partial _{t}\varphi (x,t)
\end{equation}
commutes with \( T \). The eigenvectors of \( T \) are the ``Bloch waves'': 
\begin{equation}
\label{theta_vacua}
\left| \vartheta _{n}\right\rangle =\frac{1}{\sqrt{k}}\sum ^{k-1}_{m=0}e^{im\vartheta _{n}}\left| m\right\rangle \: ,\qquad \vartheta _{n}=\frac{2\pi }{k}n\: ,
\end{equation}
\begin{equation}
\label{T_theta_vacua_action}
T\left| \vartheta _{n}\right\rangle =e^{-i\vartheta _{n}}\left| \vartheta _{n}\right\rangle \,,
\end{equation}
which are eigenstates of the Hamiltonian as well. In fact, these states can
be continued to finite volume and are eigenstates of \( H \) when \( L<\infty  \),
while there are no natural counterparts of the states \( \left|
n\right\rangle  \): for finite \( L \) we define them using the inverse formula
\[
\left| m\right\rangle =\frac{1}{\sqrt{k}}\sum ^{k-1}_{n=0}e^{-im\vartheta _{n}}\left| \vartheta _{n}\right\rangle \: .\]
 One can also introduce a \( {\mathbb Z}_{2} \) transformation \( S \) which
is defined as 
\begin{equation}
\label{S_symmetry}
S\varphi (x,t)S^{-1}=-\varphi (x,t)\: ,\qquad S^{\dagger }=S^{-1}=S\,.
\end{equation}
It acts on the ground states as
\begin{equation}
\label{S_ground_state_action}
S\left| n\right\rangle =\left| k-n\right\rangle 
=\left| -n\right\rangle \: ,\qquad S\left| \vartheta _{l}\right\rangle 
  =\left| -\vartheta _{l}\right\rangle =\left| \vartheta _{k-l}\right\rangle\,,
\end{equation}
and commutes with the Hamiltonian \( H \). The \( S \) and \( T \) transformations
together generate the discrete group \( {\mathbb D}_{k} \).

\subsection{The spectrum of local operators \label{subsec:local_operators}}

As a consequence of (\ref{period}, \ref{quasi_periodic}), the
exponential fields\footnote{In writing $V_{m}$ we assume that the
exponential fields are normal ordered and are normalized such that the short
distance limit of their two point function is
\[\langle V_{m}(x)V_{-m}(y)\rangle \rightarrow 
\vert x-y\vert^{-\frac{\beta^2m^2}{k^24\pi}}\quad {\rm for}\quad \vert
x-y\vert\rightarrow 0\,.\]}
\begin{equation}
\label{electric_vertex_ops}
V_{m}=\exp \left( {i\beta }\frac{m}{k}\varphi \right) 
\end{equation}
are well-defined local operators  
provided \( m\in {\mathbb Z} \).
One can easily compute
\begin{equation}
\label{ST_action_on_vertex_ops}
TV_{m}T^{-1}=e^{-\frac{2\pi i}{k}m}V_{m}\; ,\qquad SV_{m}S^{-1}=V_{-m}\,.
\end{equation}
In the short distance limit, the behaviour of the correlation functions is described
by a \( c=1 \) compactified free boson with the Lagrangian density
\begin{equation}
\label{UV_boson}
{\cal L}=\frac{1}{8\pi }\partial _{\mu }\chi \partial ^{\mu }\chi\,.
\end{equation}
 In order to determine the complete spectrum of local operators, let us consider
this UV limiting theory. The field \( \chi  \) lives on a circle with compactification
radius \( r \)
\[
\chi \sim \chi +2\pi nr\: ,\qquad n\in {\mathbb Z}\]
 where \( r \) can be determined from (\ref{period}). Taking into account
that the normalizations of \( \varphi  \) and \( \chi  \) differ by a factor
of \( \sqrt{4\pi } \) we get
\begin{equation}
\label{compactification_radius}
r=\frac{\sqrt{4\pi }}{\beta }k\, .
\end{equation}
The above theory has a \( \widehat{U(1)}_{L}\times \widehat{U(1)}_{R} \) Kac-Moody
symmetry. The primary fields under this algebra are vertex operators \( {\cal V}_{(n,w)} \)
with left/right conformal weights given by
\begin{equation}
\label{vertex_ops_dim}
\Delta _{(n,w)}^{\pm }=\frac{1}{2}\left( \frac{n}{r}\pm \frac{wr}{2}\right) ^{2}\, \, .
\end{equation}
\( n \) is the field momentum quantum number, while \( w \) is the so-called
winding number. The local operators \( V_{m} \) correspond to \( {\cal V}_{(m,0)} \).
The sine--Gordon potential can be identified as 
\begin{equation}
\label{perturbing_op_ident}
\nord{\cos \beta \varphi }\equiv \frac{1}{2}(\mathcal{V}_{(k,0)}+\mathcal{V}_{(-k,0)})\, \, .
\end{equation}
 There are only two possible maximal local operator algebras in a \( c=1 \)
free boson theory with compactification radius \( r \) \cite{heretic}: 
\begin{eqnarray}
\label{local_algebras}
{\cal A}_{b}&=&\{\mathcal{V}_{(n,w)}:\, \, w\in \mathbb {Z},\, n\in
\mathbb {Z}\}\,,
\nonumber\\
{\cal A}_{f}&=&\{\mathcal{V}_{(n,w)}:\, \, w\in \mathbb {Z},\, n\in
\mathbb {Z}+w/2\}\,.
\end{eqnarray}
The first one corresponds to a bosonic model, while for the second one the operators
corresponding to \( w \) odd are fermions. This
gives the complete list of possible local theories in the ``sine--Gordon class''
where standard sine--Gordon/massive Thirring corresponds to choosing \( k=1 \)
and the algebra \( {\cal A}_{b} \)/\( {\cal A}_{f} \) , respectively.

\subsection{Truncated Conformal Space for \protect\( \mathrm{SG}(\beta ,k)\protect \) \label{subsec:tcsa_description}}

In order to support our theoretical considerations we shall use the
well-known Truncated Conformal Space Approach (TCSA) to obtain
numerical data for the model \( \mathrm{SG}(\beta ,k) \).  This method
was developed by Yurov and Zamolodchikov \cite{yz_tcs} for
perturbations of Virasoro minimal models and more recently extended to
perturbations of \( c=1 \) theories in \cite{frt_excited} where we
refer the interested reader for more detail.

Here we limit ourselves to recalling the basic facts. We can represent
the Hamiltonian (\ref{sG_Hamiltonian}) as an infinite hermitian matrix
on the space of states of the \( c=1 \) conformal field theory built
on the algebra \( {\cal A}_{b} \) (\ref{local_algebras})\footnote{From
now on we will restrict ourselves to the model based on the algebra
${\cal A}_b$, as its fermionic counterpart is very similar.}:
\begin{equation}
\label{UV_Hilbert_space}
{\cal H}=\bigoplus _{{\cal V}_{(n,m)}\in {\cal A}_{b}}{\cal F}_{(n,m)}\,,
\end{equation}
where 
\begin{equation}
\label{Fock_space}
{\cal F}_{(n,m)}=\mathrm{span}\, \left\{ a_{-k_{1}}\ldots
a_{-k_{l}}\left| {\cal V}_{(n,m)}\right\rangle \: ,\: k_{1},\ldots ,k_{l}\in {\mathbb Z}_{+}\right\} 
\end{equation}
is the Fock space built over the primary state \( \left| {\cal V}_{(n,m)}\right\rangle  \)
with the negative frequency modes of the conformal free boson field \( \chi  \)
(\ref{UV_boson}). The Hamiltonian takes the form
\begin{equation}
\label{TCS_Hamiltonian}
H=\frac{2\pi }{L}\left( L_{0}+\bar{L}_{0}-\frac{c}{12}{\mathrm{Id}}+\lambda \frac{L^{2-h}}{\left( 2\pi \right) ^{1-h}}B\right) \: ,
\end{equation}
where \( L_{0} \) and \( \bar{L}_{0} \) are diagonal matrices with their diagonal
elements being the left and right conformal weights, \( {\mathrm{Id}} \) is the identity
matrix, 
\begin{equation}
\label{potential_UV_dim}
h=\frac{\beta ^{2}}{4\pi }=\frac{2p}{p+1}
\end{equation}
is the scaling dimension of the perturbing potential and the matrix elements
of \( B \) between two states \( \left| \Phi \right\rangle  \) and \(
\left| \Psi \right\rangle  \) are
\begin{equation}
\label{B_matrix}
B_{\Phi ,\Psi }=\frac{1}{2}\left\langle \Phi \right|{\cal
V}_{(k,0)}(1,1)+{\cal V}_{(-k,0)}(1,1)\left| \Psi \right\rangle \: .
\end{equation}
The parameter $\lambda$ is connected to $\mu_0$ in
eqn.~(\ref{sG_action}) via a known relation \cite{mass_scale}.  The
matrix elements of \( B \) can be calculated in closed form. We would
like to call attention to the fact that the matrix elements of all
vertex operators \( {\cal V}_{(n,m)} \) in the basis
(\ref{Fock_space}) are real numbers.  As a result the Hamiltonian is a
real symmetric matrix which will be important in establishing
(\ref{vev_reality}).

We choose our units in terms of the soliton mass \( M \) which is related to
the coupling constant \( \lambda  \) by the mass gap formula obtained from
TBA in \cite{mass_scale}: 

\begin{equation}
\label{mass_gap}
\lambda =\kappa(h) M^{2-h},
\end{equation}
where 
\begin{equation}
\label{kappa}
\kappa(h) =\frac{2\Gamma (h/2)}{\pi \Gamma (1-h/2)}\left( \frac{\sqrt{\pi}
\Gamma \left( \frac{1}{2-h}\right)
}{2 \Gamma \left( \frac{h}{4-2h}\right) }\right) ^{2-h}\, .
\end{equation}
(Note that we use the same massgap relation in the $k$-folded model as
in ordinary sine--Gordon, due to our previous argument about the
relation between local properties in the two models). In
what follows we normalize the energy scale by taking \( M=1 \) and
denote the dimensionless volume \( ML \) by \( l \). For numerical
computations, we shall use the dimensionless Hamiltonian

\begin{equation}
\label{dimlessham}
{\hat H}=\frac{H}{M}=\frac{2\pi }{l}\left( H_{\mathrm{CFT}}+\kappa(h)
\frac{l^{2-h}}{\left( 2\pi \right) ^{1-h}}B\right)\, ,\qquad H_{\mathrm{CFT}}=L_{0}+\bar{L}_{0}-\frac{c}{12}{\mathrm{Id}}\: .
\end{equation}
We diagonalize the matrix \( {\hat H} \) in a truncated Hilbert space defined as 
\begin{equation}
{\cal H}_{\mathrm{TCS}}(s,w,E_{\mathrm{cut}})=\left\{ \left| \Psi \right\rangle :\: \left( L_{0}-\bar{L}_{0}\right) \left| \Psi \right\rangle =s\left| \Psi \right\rangle ,\: Q_{k}\left| \Psi \right\rangle =w\left| \Psi \right\rangle ,H_{\mathrm{CFT}}\left| \Psi \right\rangle \leq E_{\mathrm{cut}}\left| \Psi \right\rangle \right\} \: ,
\end{equation}
where, besides imposing an upper bound on conformal energy, we restricted the Hilbert space
to a given value of the conformal spin \( L_{0}-\bar{L}_{0} \) and of the topological
charge \( Q_{k} \) (the winding number of the free boson \( \chi  \)) since
these commute with the Hamiltonian \( H \). As in the repulsive regime
($p>1$) the TCSA for the SG theory is plagued by UV problems
\cite{frt_excited}, in this paper we restrict all of our numerical
studies to the attractive regime ($p<1$). For the purposes of
producing the numerical data used later in the paper we typically used
values of $E_{\mathrm{cut}}$ that correspond to roughly $6000$--$12000$ states.

\section{\protect\( \mathrm{SG}(\beta ,k)\protect \) in the NLIE framework \label{sec:NLIE_description}}

In a finite spatial volume the spectrum (the ground and the excited
states) of sine--Gordon theory (\ref{sG_action}) is described by a
nonlinear integral equation (NLIE) \cite{kbp, ddv}. We shall start
with a more general equation than in the normal sine--Gordon situation
by introducing a twist angle $\vartheta$ \'{a} la Zamolodchikov
\cite{polymer}, originally motivated by considerations related to
polymers. Later it appeared in the description of the finite volume
spectrum of Virasoro minimal models perturbed by $\Phi_{(1,3)}$
\cite{frt_alpha}. It corresponds to switching on a chemical potential
coupled to the topological charge \cite{fs}. The full twisted NLIE reads
\begin{eqnarray}
Z(\lambda )=ML\sinh \lambda +g(\lambda |\lambda _{j})+\vartheta -i\int
^{\infty }_{-\infty }\ud x\: G(\lambda -x-i\eta )\log \left( 1+(-1)^{\delta }e^{iZ(x+i\eta )}\right)  &  & \nonumber \\
+i\int ^{\infty }_{-\infty }\ud x\: G(\lambda -x+i\eta )\log \left( 1+(-1)^{\delta }e^{-iZ(x-i\eta )}\right)  &  &\,, \label{alpha-nlie-excited} 
\end{eqnarray}
where \( M \) is the soliton mass, \( L \) is the volume, \( \eta \)
is a suitably chosen real shift, $\delta$ is either $0$ or $1$, and
the kernel \( G \) reads
\begin{equation}
\label{ddv_kernel}
G(\lambda )=\frac{1}{2\pi }\int ^{+\infty }_{-\infty }\ud k\, e^{ik\lambda }\frac{\sinh \frac{\pi (p-1)k}{2}}{2\sinh \frac{\pi pk}{2}\, \cosh \frac{\pi k}{2}}\, .
\end{equation}
The twist angle \( \vartheta  \) can be restricted to lie in the range \( -\pi <\vartheta \leq \pi  \)
without loss of generality: this choice simplifies the description of
the results of the UV calculations. 

The term \( g(\lambda |\lambda _{j}) \) is the so-called source term, composed
of contributions from the holes, special objects (roots/holes) and complex roots.
The form of \( g \) is specific to the state in the spectrum one wants to describe;
for states with no particles, \( g=0 \) (at least for \( L \) large enough,
where the so-called special objects do not appear). We denote their positions
by the general symbol \( \{\lambda _{j}\}=\{h_{k}\, ,\, y_{k}\, ,\, c_{k}\, ,\, w_{k}\} \)
(\( h \) stands for holes, \( y \) for special objects and \( c \) (\( w \))
for close (wide) complex roots). The source term takes the general form
\[
\displaystyle g(\lambda |\lambda _{j})=\sum ^{N_{H}}_{k=1}\chi (\lambda -h_{k})-2\sum ^{N_{S}}_{k=1}\chi (\lambda -y_{k})-\sum ^{M_{C}}_{k=1}\chi (\lambda -c_{k})-\sum ^{M_{W}}_{k=1}\chi (\lambda -w_{k})_{\mathrm{II}}\: ,\]
where 
\begin{equation}
\label{chi}
\chi (\lambda )=2\pi \int _{0}^{\lambda }\ud x\,G(x)
\end{equation}
and the second determination \emph{}for any function \( f(\lambda ) \)
is defined by
\begin{equation}
\label{2nd_determination}
f(\lambda )_{\mathrm{II}}=\left\{ \begin{array}{ll}
f(\lambda )+f\left( \lambda -i\pi \mathrm{sign}\left( \Im m\lambda \right) \right) \: , & p>1\: ,\\
f(\lambda )-f\left( \lambda -i\pi \mathrm{sign}\left( \Im m\lambda \right)p \right) \: , & p<1\: ,
\end{array}\right. 
\end{equation}
whenever \( |\Im m\lambda |>\min (\pi ,\pi p) \).

The source positions \( \lambda _{i} \) are determined from the Bethe quantization
conditions
\begin{equation}
\label{quantum}
Z(\lambda _{j})=2\pi I_{j}\, \, ,\qquad I_{j}\in {\mathbb Z}+\frac{1-\delta }{2}\, \, ,
\end{equation}
where \( I_{j} \) are the Bethe quantum numbers (for wide roots one must use
the second determination of \( Z \), defined as in (\ref{2nd_determination})). 

Given a solution for $Z$, the energy and momentum of the state can be
computed from the formulae
\begin{eqnarray*}
E & = & E_{\mathrm{bulk}}+M\sum ^{N_{H}}_{j=1}\cosh h_{j}-2M\sum ^{N_{S}}_{j=1}\cosh y_{j}\\
 &  & -M\sum ^{M_{C}}_{j=1}\cosh c_{j}-M\sum _{j=1}^{M_{W}}(\cosh w_{j})_{\mathrm{II}}\\
 &  & -M\int ^{\infty }_{-\infty }\frac{\ud x}{2\pi }2\Im m\left[ \sinh (x+i\eta )\log (1+(-1)^{\delta }e^{iZ(x+i\eta )})\right] \: ,
\end{eqnarray*}
 
\begin{eqnarray*}
P & = & M\sum ^{N_{H}}_{j=1}\sinh h_{j}-2M\sum ^{N_{S}}_{j=1}\sinh y_{j}\\
 &  & -M\sum ^{M_{C}}_{j=1}\sinh c_{j}-M\sum _{j=1}^{M_{W}}(\sinh w_{j})_{\mathrm{II}}\\
 &  & -M\int ^{\infty }_{-\infty }\frac{\ud x}{2\pi }2\Im m\left[ \cosh (x+i\eta )\log (1+(-1)^{\delta }e^{iZ(x+i\eta )})\right] \: ,
\end{eqnarray*}
 where
\begin{equation}
\label{univ_bulk_energy}
E_{\mathrm{bulk}}=-\frac{1}{4}M^{2}L\tan \frac{\pi p}{2}\: .
\end{equation}
 For more details on the NLIE for excited states we refer to the literature
\cite{frt_excited, frt_alpha, fmqr_excited, ddv_excited}. 

A detailed calculation of the UV limit of the twisted NLIE (\ref{alpha-nlie-excited})
was performed in \cite{frt_alpha}. For our case, remembering that the identification
of the perturbing potential (\ref{perturbing_op_ident}) is different from that
of Feverati et al., we have to trace the appearance of the sine--Gordon twist
parameter \( k \) in their formulas. Provided one chooses 
\begin{equation}
\label{sG_NLIE_quant_rule}
\delta =N_{H}\, \bmod \, 2\: ,
\end{equation}
the ultraviolet limit is described by states in the conformal family of vertex
operators of the form
\[
\mathcal{V}_{\left( n,m\right) }\, ,\qquad n\in k{\mathbb
Z}+\frac{k\vartheta }{2\pi }\,,
\qquad m=\frac{1}{k}\left( N_{H}-2N_{S}-M_{C}-2M_{W}\theta (p-1)\right)
\,,\]
where $\theta(x)$ denotes the Heaviside step function.

Therefore to describe the UV spectrum \( {\cal A}_{b} \) (\ref{local_algebras})
of the \( k \)-folded sine--Gordon theory one must choose the following values
for the twist parameter:
\begin{equation}
\label{twist_values}
\vartheta =\frac{2\pi n}{k}\,,\qquad n=\left[ -\frac{k}{2}+1\right] ,\ldots ,\left[ \frac{k}{2}\right] 
\end{equation}
which is $\bmod\,2\pi$ equivalent to the set of $\vartheta_n$ values in
(\ref{theta_vacua}).

We also have to restrict
\begin{equation}
\label{NLIE_topcharge_restriction}
{\cal Q}=N_{H}-2N_{S}-M_{C}-2M_{W}\theta (p-1)\in k{\mathbb Z}\: .
\end{equation}
The quantity $\cal Q$ counts the topological charge in the units of
the usual sine--Gordon theory ($\mathrm{SG}(\beta,k=1)$). The formula
(\ref{NLIE_topcharge_restriction}) expresses the fact that all states of
$\mathrm{SG}(\beta,k)$ which satisfy the periodic boundary conditions must
be compatible with (\ref{quasi_periodic}) as a consequence.

\section{Vacuum structure and instantons in finite volume \label{sec:instantons}}

In this section we investigate the vacuum structure of the $k$-folded
sine--Gordon model and show that it provides a laboratory to further
test the NLIE as well as for analyzing the higher order corrections to
the dilute instanton gas approximation (DIG). We test the NLIE by
comparing its predictions both to the results of the instanton
calculus and to the TCSA data.  The second possibility is due to the
fact that terms beyond the DIG give contributions to certain
quantities, where they are not suppressed by lower order terms. First
we give a group theoretical, thus qualitative description of the
vacuum spectrum, then we analyse it quantitatively using the NLIE and
finally we make the instanton calculation.

\subsection{Symmetry considerations}

In analyzing the vacuum structure we start with the \( L\to \infty  \) limit.
In this case Derrick's theorem forbids the existence of instantons. 
We have \( k \) different degenerate minima with 
the corresponding states given by $\vert m\rangle $, $ m=0,\dots,k-1$. 
The symmetry transformations act on $\vert m\rangle $ by
eqns.~(\ref{Tdef}), (\ref{S_ground_state_action}) and they commute with
the diagonal Hamiltonian.

Now we consider the theory for finite \( L \).  For finite \( L \)
instantons exist and they lift the degeneracy of the ground states. As
a consequence we have a unique vacuum state invariant under \(
{\mathbb D}_{k} \) and the Hamiltonian is no longer diagonal in the
basis spanned by \( \vert {m}\rangle \). Nevertheless from its
symmetry properties we can determine its form.  Since it commutes with
the symmetry transformations it must belong to the center of the group
algebra of \( {\mathbb D}_{k} \). The generators of the center can be
obtained by summing up all elements of a given conjugacy class with
the same weights. One can show by direct computation that in the
particular representation considered the elements of the center which
correspond to the $S$ transformation can be expanded in terms of those
generators which correspond purely to the $T$
transformation. Consequently, the most general element of the center
has the following form:
\[
H=E_{0}(L){\rm Id}+\tilde{E}_{1}(L)(T+T^{-1})+\dots +\tilde{E}_{i}(L)(T^{i}+T^{-i})+\dots \:.\]
We can diagonalize this matrix by diagonalizing its commutant, i.e.~the representation
matrices of $T$ itself. The eigenvectors are the states
$\vert\vartheta_m\rangle $ introduced in eqn.~(\ref{theta_vacua}) and  
the corresponding eigenvalues of the Hamiltonian are:
\begin{equation}
\label{espt}
H_{m}=E_{0}(L)+\sum\limits_{j=1}^{[k/2]}E_j(L)\cos\left(\frac{2\pi}{k}jm\right)\,,\quad
m=0,\dots ,k-1\,,\quad 2\tilde{E}_j=E_j\,.
\end{equation}
Thus for finite $L$, instead of the $k$-fold degenerate ground states
we have the following vacuum structure: there is a nondegenerate
lowest eigenvalue, $H_0$, while the rest of the eigenvalues come in
pairs, $H_r=H_{k-r}$ for $1\leq r\leq [k/2]$, at least for odd $k$.
For $k$ even $H_{k/2}$ is non-degenerate as well. Clearly from the
knowledge of the energy levels \( H_{m} \) we can recover all the
coefficients \( E_{j}(L) \).

\subsection{Leading finite size corrections to the vacuum energy \label{subsec:nlie_calculation}}

The NLIE for the twisted vacuum reads
\begin{eqnarray}
Z(\lambda )=ML\sinh \lambda +\vartheta &-&i\int ^{\infty }_{-\infty }\ud
x\: G(\lambda -x-i\eta )\log \left( 1+e^{iZ(x+i\eta )}\right)  \nonumber \\
&+&i\int ^{\infty }_{-\infty }\ud x\: G(\lambda -x+i\eta )\log \left( 1+e^{-iZ(x-i\eta )}\right) \, \, , \label{twisted-nlie} 
\end{eqnarray}
Here we relax the condition (\ref{twist_values}) on \( \vartheta \) in
order to keep the calculation more general. As \( \vartheta \,
\rightarrow \, \vartheta +2\pi \) is a symmetry of the NLIE
(\ref{twisted-nlie}) we can restrict the value of the twist angle \(
\vartheta \) to \( -\pi <\vartheta \leq \pi \). The value \( \vartheta
=0 \) corresponds to the untwisted sector. The vacuum energy can be
obtained using
\begin{equation}
\label{ddv_energy}
E(L)=-2M\Im m\int ^{\infty }_{-\infty }\frac{\ud x}{2\pi }\sinh (x+i\eta )\log \left( 1+e^{iZ(x+i\eta )}\right) \, \, ,
\end{equation}
where the counting function \( Z(x) \) is a solution of the vacuum
NLIE (\ref{twisted-nlie}) and we omitted the universal bulk energy
term (\ref{univ_bulk_energy}). For large values of the volume \( L \),
the solution is
\begin{equation}
\label{zeroth_order_Z}
Z(x)\approx \vartheta +l\sinh (x)\, \, ,\, \, l=ML\, \, .
\end{equation}
The value of \( \eta  \) must lie inside the analyticity strip
for the kernel \( G \), i.e.~\( |\eta |<\min (\pi ,\pi p) \) (``first determination'').
However, there is no singularity whatsoever in the integrand of \( E(L) \)
when using the above approximation for \( Z \) and the contour can be shifted
to \( \eta =i\pi /2 \) for any value of \( p \) (in fact, the result
is independent of \( p \) itself). One obtains
\begin{equation}
\label{free_fermi_gas}
E(\vartheta , L)=-M\int ^{\infty }_{-\infty }\frac{\ud x}{2\pi }\cosh (x)\left[ \log \left( 1+e^{i\vartheta -l\cosh x}\right) +\log \left( 1+e^{-i\vartheta -l\cosh x}\right) \right] \, .
\end{equation}
 Expanding the term \( \log (1+\ldots ) \) in Taylor series, a short
 computation gives
\begin{equation}
\label{free_fermi_exp}
E(\vartheta , L)=-\frac{2M}{\pi }\sum ^{\infty }_{n=1}\frac{(-1)^{n-1}}{n}K_{1}(nl)\cos n\vartheta \, \, ,
\end{equation}
where \( K_{1}(z) \) is a modified Bessel function of the second kind:
\[
K_{1}(z)=\int ^{\infty }_{-\infty }\frac{\ud x}{2}\cosh (x)e^{-z\cosh x}\, \, .\]
We remark that this result is exact at the free fermion point \( p=1 \) (where
\( G=0 \) due to (\ref{ddv_kernel})) and it can be shown to resum into
\begin{eqnarray}
E(\vartheta , L) & = & -\frac{\pi }{6L}\left\{ 1-3\left( \frac{\vartheta }{\pi }\right) ^{2}-\frac{3l^{2}}{2\pi ^{2}}\left( 1-2C-2\log \frac{l}{\pi }\right) +\right. \nonumber \\
 &  & \frac{6}{\pi }\sum ^{\infty }_{m=1}\left( \sqrt{((2m-1)\pi +\vartheta )^{2}+l^{2}}-(2m-1)\pi -\frac{l^{2}}{2(2m-1)\pi }\right) +\nonumber \\
 &  & \left. \frac{6}{\pi }\sum ^{\infty }_{m=1}\left( \sqrt{((2m-1)\pi -\vartheta )^{2}+l^{2}}-(2m-1)\pi -\frac{l^{2}}{2(2m-1)\pi }\right) \right\} \label{free_fermion_alpha} 
\end{eqnarray}
with \( C=0.57721566\ldots  \) being the Euler-Mascheroni constant. This is exactly
the result for a free Dirac fermion with twisted boundary conditions in finite
volume \( L \).

\subsection{Instantons in finite volume \label{subsec:instanton_interpretations}}

With the exception of the leading \( n=1 \) term in (\ref{free_fermi_exp}), all the others get further corrections from
the integral term in the NLIE. Therefore for a general value of \( p \) the
series (\ref{free_fermi_exp}) must be truncated to its first term for consistency.
Using the asymptotic behaviour of \( K_{1}(z) \) we obtain 
\begin{equation}
\label{dilute_gas}
\frac{E(\vartheta , L)}{M}=-\sqrt{\frac{2}{\pi l}}e^{-l}\cos \vartheta +\ldots \, \, .
\end{equation}
 
\begin{figure}
{\par\centering \includegraphics{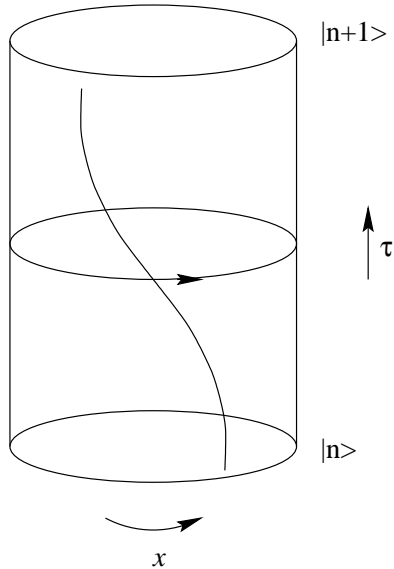} \par}

\caption{The one-instanton contribution to the vacuum energy}\label{instspace}
\end{figure}
\noindent When the twist angle takes the allowed values (\ref{twist_values})  
\(
\vartheta _{m}=2\pi m/k\), this result can be compared to
(\ref{espt}), showing that the leading \( n=1 \) term in
(\ref{free_fermi_exp}) predicts the value of $E_1(L)/M$.
 
It is easy to interpret this result in terms of the usual instanton
calculus.  The one-instanton configuration is nothing else but the
static one-soliton configuration of sine--Gordon theory on a cylinder
replacing the space variable by the Euclidean time. It is independent
of the spatial coordinate and so satisfies the periodic boundary
condition on the cylinder. The Euclidean action is just the soliton
mass \( M \) multiplied by the volume \( L \). This gives us the
factor \( e^{-l} \) by the normal rules of instanton calculus. The
factor \( \sqrt{2/(\pi l)} \) is composed of two parts: a
contribution of \( \sqrt{l/(2\pi )} \) comes from the one
(bosonic) zero mode of the instanton generated by translations and the
rest from the determinant of the nonzero mode oscillations around the
soliton, truncated to quadratic terms in the action. One then gets the
result (\ref{dilute_gas}) from the usual dilute instanton gas
calculation (cf.~(\ref{dilute_gas_result})).  (Since this determinant
is different from the one needed to compute the quantum corrections to
the SG soliton's mass we spell out the details of the calculation in
appendix \ref{app:instanton_calculus}).  Thus for $E_1(L)/M$ the $n=1$
term of the NLIE and the DIG give identical predictions. This is
interesting, as the former one is expected to give reliable results
for $p\sim 1$, while of the latter we expect this for $p\sim 0$;
however, as we have seen the leading term is independent of $p$. Note
that this prediction is also independent of the folding number $k$.
\begin{figure}
{\par\centering \includegraphics{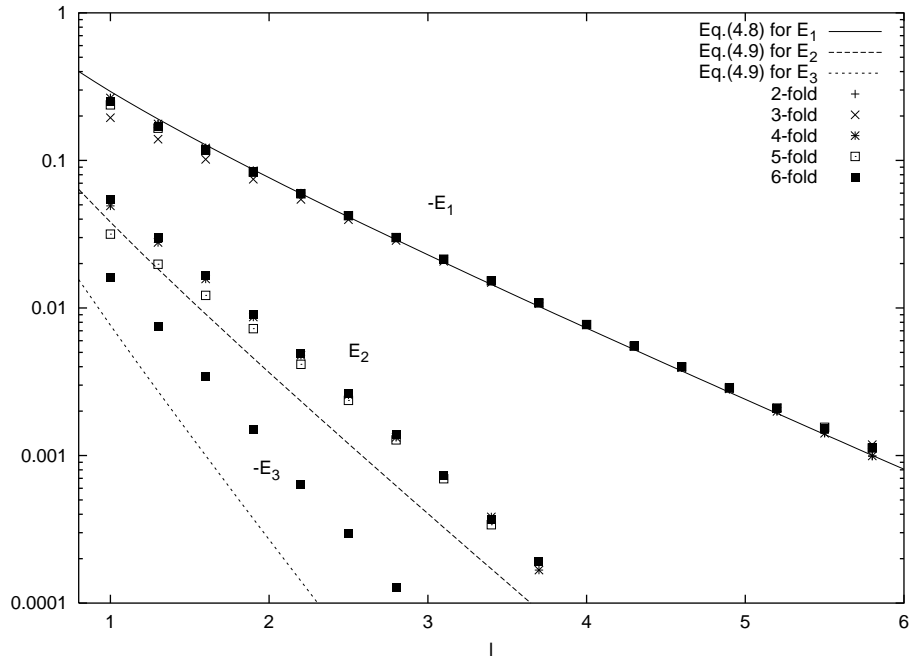} \par}

\caption{TCSA data versus predictions for \( E_1\), \( E_2\) and \(
E_3\) at \( p=2/7\) }\label{kdep}
\end{figure}

\begin{figure}
{\par\centering \includegraphics{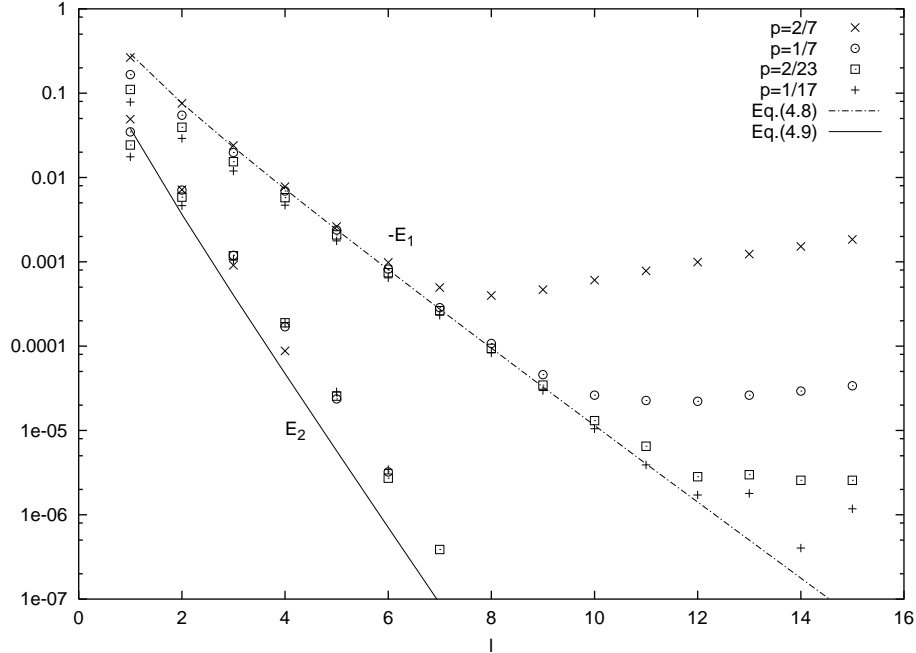} \par}
\caption{The \( p\) (in)dependence of \( E_1\) and \( E_2\) in various
4-folded models}\label{pdep}
\end{figure}

To obtain a theoretical prediction for $E_j(L)$, $j>1$ in (\ref{espt}),
one must go beyond these approximations. In the NLIE---as mentioned
earlier---this would necessitate the inclusion of the integral term,
while in the instanton calculus it would require a handle on the
\emph{exact} multi-instanton solutions (as opposed to the \emph{approximate}
ones in DIG) together with their determinants 
in the cylindrical spacetime on Fig.~\ref{instspace}. To derive an
explicit expression for  $E_j(L)$ is beyond the scope of the present
paper, and we merely note that we expect
\begin{equation}\label{e2est}
\frac{E_j(L)}{M}=C(j,p,k)\frac{e^{-jl}}{l^{m(j,p,k)}}\left( 
1+{\mathcal{O}}(e^{-(j+k)l})\right)\,.\end{equation}
Keeping simply the first few terms with higher \( n\) in (\ref{free_fermi_exp})
predicts 
\[ C(j,p,k)=(-1)^j\sqrt{\frac{2}{\pi j^3}}\,,\]
and \(m=1/2\) for \( j=2,3,\dots,\) and on Fig.~\ref{kdep} and
Fig.~\ref{pdep} we used these values for making predictions about \(
E_2\) and \( E_3\).  Since the NLIE is an exact description, the
constants $C(j,p,k)$ and the exponents $m$ can in principle be
determined exactly (or numerically with very high accuracy).

To test the NLIE and instanton predictions we determined numerically
$E_1$, $E_2$ and $E_3$ from the first different $[k/2]$ TCSA
eigenvalues using eqn.~(\ref{espt}). On Fig.~\ref{kdep} we collected
these quantities for models having the same $p=2/7$ but differing in
their folding number, which varied between 2 and 6. The data show a
universal behaviour with no folding number dependence and in case of
$E_1$ they fit very well to the NLIE/instanton prediction,
eqn.~(\ref{dilute_gas}). Please note that the data and the most naive
NLIE prediction (\ref{e2est}) for $E_2$ and $E_3$ differ only in
the prefactor $C(j,p,k)$ as the predictions run parallel to the data
in the semilogarithmic plot. On Fig.~\ref{pdep} we compile the
numerical values of $E_1$ and $E_2$ in 4-folded models having
different values of $p$. Again the data show a universal behaviour with no
significant $p$ dependence, and the prediction (\ref{dilute_gas}) for
$E_1$ describes the data very well. This figure also shows that the
smaller $p$ is the larger is the $l$ range where TCSA gives reliable
data.

\section{Multi-particle energy levels in finite volume \label{section:multi_particle_levels}}

The finite volume spectra of completely integrable models can be used
to test their conjectured exact \( S \)-matrices. In particular, the
energy levels of multi-kink states satisfying the periodic boundary
conditions can be determined in terms of their \( S \)-matrices
\cite{KM3}. These energy levels then can be compared to the finite
volume spectrum obtained by TCSA.

There are three physical effects that contribute to the finite volume
energy levels of a QFT. The `tunnelling' effects are there in any
theory---like \( \mathrm{SG}(\beta ,k) \)---which has degenerate vacua in
infinite volume. The corrections due to tunnelling are \(
{\mathcal{O}}(\exp (-ML)) \), where \( L \) is the volume of the
compact coordinate and \( M \) is a characteristic mass, in our case
the mass of the quantum kink. In any massive theory there are two
types of `off-shell' effects due to the vacuum polarization and the
interactions mediated by virtual particles, but both of them give \(
{\mathcal{O}}(\exp (-ML)) \) corrections only. Finally in finite
volume the particles continuously scatter on each other in a
multi-particle state and as a result of these `scattering' effects the
stationary scattering states (i.e.~the ones invariant under the mutual
scattering of the constituents) are the true energy eigenstates. The
resulting quantization conditions---called Bethe--Yang equations in
\cite{KM3}---can be expressed in terms of the \( S \)-matrix if we
assume that the particles are point-like. The corresponding corrections
to the energy levels are \( {\mathcal{O}}(L^{-2}) \) and so much
larger than in the previous two cases, therefore they provide an ideal
possibility to check the $S$-matrices.

\subsection{Particle spectrum in the classical \protect\(
\mathrm{SG}(\beta ,k)\protect \) in infinite
volume\label{subsec:classical_particle_spectrum}}

As mentioned above the static soliton solution of the SG theory
becomes a kink in the $k$-folded model.
 More precisely the classical kink
solutions of \( \mathrm{SG}(\beta ,k) \), connecting neighbouring minima can be written
with the aid of the SG soliton solution as 
\begin{equation}
\label{1kink}
K_{n,n+1}(x,t)=\frac{4}{\beta }\arctan e^{\mu
_{0}(x-x_{0})}+\frac{2n\pi }{\beta }
\qquad n=0,\ldots ,k-1;\: k\equiv 0\, .
\end{equation}
 Thus instead of the single soliton we have \( k \) different `one
kink' solutions in \( \mathrm{SG}(\beta ,k) \). (The antikink
solutions are obtained by the \( \varphi \mapsto -\varphi \)
reflection). Since a multi-kink solution corresponds to a sequence of
vacua on a line one cannot arbitrarily compose single kinks to obtain
an allowed solution, in contrast to the SG (anti)solitons. This
restriction on the sequence of kinks translates into restrictions on
the multi-particle Hilbert space in the quantized model. Note however
that any multi-soliton/antisoliton solution of SG has a (multi-)kink
interpretation in \( \mathrm{SG}(\beta ,k) \), at least in infinite
volume with no boundary conditions prescribed.

\( \mathrm{SG}(\beta ,k) \) also has \( k \) different breather
solutions which oscillate around the \( k \) different minima of the
potential:
\begin{equation}
\label{bre}
B^{(v)}_{n}(x,t)=\frac{4}{\beta }\arctan \frac{\sin \left( \frac{\mu
_{0}vt}{\sqrt{1+v^{2}}}\right) }{v\cosh \left( \frac{\mu
_{0}x}{\sqrt{1+v^{2}}}\right) }+\frac{2n\pi }{\beta }\: ,\quad v\in
{\mathbb R}\: ,\quad n=0,\ldots ,k-1\: .
\end{equation}
 An important characteristic of the SG solutions is their topological charge.
It measures how many times the \( \varphi (t,x) \) field winds around its range
as \( x \) runs from \( -\infty  \) to \( \infty  \). Thus in \( \mathrm{SG}(\beta ,k) \)
we define it as 
\begin{equation}
\label{topdef}
Q_{k}=\frac{\beta }{2k\pi }\int ^{\infty }_{-\infty }\partial
_{x}\varphi \ud x\: .
\end{equation}
 This implies that the single kink solutions have a fractional \( 1/k
\) topological charge and to make \( Q_k \) integer we have to consider
at least a \( k \)-kink solution. The fractionally charged
configurations give rise to nonlocal states at the quantum level,
which explains the restriction (\ref{NLIE_topcharge_restriction})
imposed on the NLIE sources. The quasi-periodic boundary condition
(\ref{quasi_periodic}) excludes all the single kink solutions, but the
breathers (\ref{bre}) satisfy it, as do the multi-kink ones with
integer topological charge (\ref{topdef}), at least approximately for
\( L \gg M_{\rm{class}}^{-1} \) where \( M_{\rm{class}}=8\mu_0/
\beta ^{2} \) is the classical kink mass.

\subsection{The particle spectrum of the quantum \protect\( \mathrm{SG}(\beta ,k)\protect \) \label{subsec:quantum_particle_spectrum}}

\begin{figure}
\centering
\includegraphics{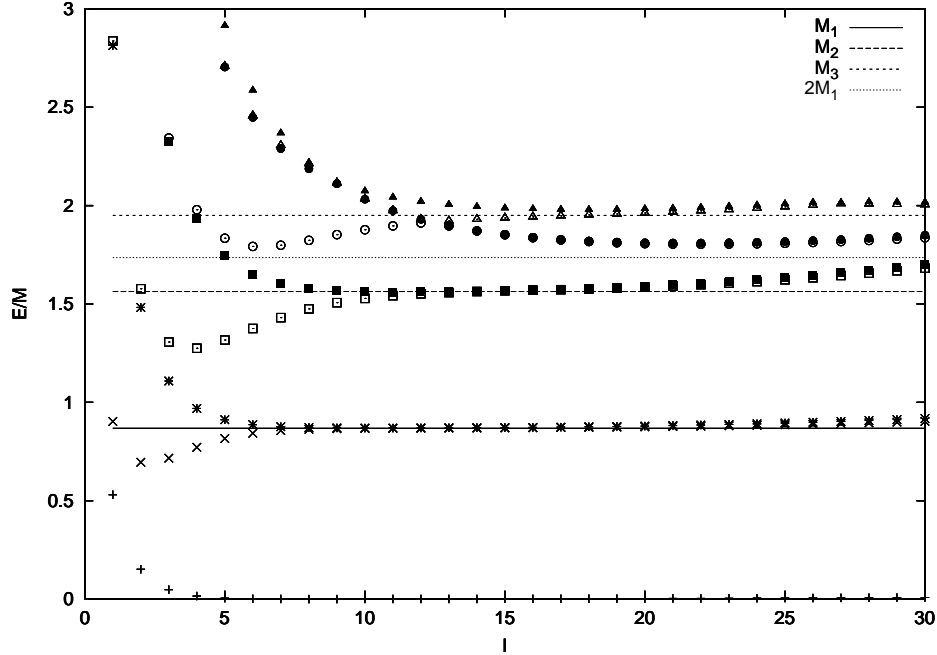}
\caption{Breather masses in the 2-folded model with \( p=2/7\)}
\label{fig:degen}
\end{figure}

In infinite volume, as a result of its integrability, the quantized SG model
contains particles corresponding to classical soliton or breather solutions.
Since integrability is a consequence of the local properties of the Lagrangian
as a function of \( \varphi  \) we expect that the quantized \( \mathrm{SG}(\beta ,k) \)
also has kink and breather particles. Of course the \( {\mathbb D}_{k} \) symmetry
requires that instead of the single quantum soliton (antisoliton) of SG in
the \( k \)-folded model there should be \( k \) quantum kinks (antikinks)
degenerate in mass. In addition we expect breather particles \( B_{n}^{(m)} \)
\( n=0,1,\dots k-1 \) corresponding to all the \( k \) different vacua. Furthermore,
contemplating e.g.~a semiclassical analysis, we expect that the relation between
the quantum kink mass \( M \) and the possible breather masses \( M_{m} \)
is given by the familiar expression
\begin{equation}
\label{breather_mass}
M_{m}=2M\sin \frac{\pi pm}{2}\: ,\quad m=1,\ldots ,\left[ \frac{1}{p}\right] \: ,
\end{equation}
 independently of the `vacuum' index \( n \) of \( B_{n}^{(m)}
\). This expectation is confirmed by Fig.~\ref{fig:degen} where, in
the 2 folded model, for $p=2/7$, we compare the breather masses
(horizontal lines) with the TCSA data.\footnote{On Fig.~5-8 $E/M$ on
the vertical axis stands for $[E(l)-E_0^v(l)]/M$, where $E_0^v(l)$ is
the vacuum energy, i.e.~the ground state energy in the $Q=0$ sector.}
These data are obtained with vanishing total momentum and vanishing
topological charge, and the various dots represent the first eight
energy eigenvalues above the ground state. The data corresponding to
single particle states (i.e.~to the first, second and third
breathers), tend much faster to their infinite volume values than the
two-particle lines, having \( 2M_1\) as their asymptotics. The reason
is that while in the single particle masses there are only \( {\cal
O}(e^{-l})\) finite size corrections, in the energy of two-particle
states there are \( {\cal O}(l^{-2})\) corrections coming from the
mutual scattering among the particles.

Denoting the kink of rapidity \( \theta  \), interpolating between the vacuum
\( \alpha  \) at \( x\rightarrow -\infty  \) and \( \beta  \) at \( x\rightarrow \infty  \)
by \( K_{\alpha \beta }(\theta ) \), we let the amplitude 
\[ S_{\alpha \beta }^{\gamma \delta }(\theta _{12}) \]
describe the process 
\[ K_{\alpha \gamma }(\theta _{1})+K_{\gamma \beta }(\theta
_{2})\rightarrow K_{\alpha \delta }(\theta _{2})+K_{\delta \beta
}(\theta _{1})\, , \] where \( \theta _{1}>\theta _{2} \) and \(
\theta _{12}=\theta _{1}-\theta _{2} \).  Labelling the vacua by \(
\alpha =0,1\dots ,k-1 \) the model \( \mathrm{SG}(\beta ,k) \) has
kinks \( K_{\alpha \beta } \) with \( |\alpha -\beta |=1 \) (or \( k-1
\)).  It is easy to describe the \( S \)-matrices of these kinks
\cite{KM3}: using the \( {\mathbb{Z}}_{k} \) symmetry as well as time
reversal and parity invariance one can show that every nonvanishing
kink-kink amplitude is equal to one of the following three amplitudes:
\( S_{\alpha \alpha +2}^{\alpha +1\alpha +1}(\theta _{12}) \), \(
S_{\alpha \alpha }^{\alpha +1\alpha +1}(\theta _{12}) \), \( S_{\alpha
\alpha }^{\alpha +1\alpha -1}(\theta _{12}) \), where \( \alpha \pm 1
\), \( \alpha +2 \) are understood mod \( k \). These (\( \alpha \)
independent) amplitudes are also independent of the global properties
of \( \varphi \) and thus they should be equal to the soliton-soliton
(\( S_{ss}(\theta ) \)), antisoliton-soliton reflection (\(
S_{R}(\theta ) \)), and antisoliton-soliton transmission (\(
S_{T}(\theta ) \)) amplitudes, respectively, of the ordinary SG
model. To compute multi-kink energy levels we need the explicit form
of \( S_{ss}(\theta ) \):
\[
S_{ss}(\theta )=-e^{i\delta (\theta )}\,,\quad \delta (\theta )=\int ^{\infty }_{0}\frac{d\omega }{\omega }\frac{\sin (\theta \omega )\sinh \left( \frac{\pi }{2}\omega (p-1)\right) }{\cosh \left( \frac{\pi }{2}\omega \right) \sinh \left( \frac{\pi }{2}\omega p\right) }\: .\]
 In a completely analogous way one obtains that the \( S_{n}^{(m)}(\theta _{12}) \)
\( S \)-matrix, which describes the possible breather-kink scatterings
\[
B^{(m)}_{n}(\theta _{1})+K_{n,n+1}(\theta _{2})\to K_{n,n+1}(\theta _{2})+B^{(m)}_{n+1}(\theta _{1})\,,\quad \theta _{1}>\theta _{2}\]
 coincides with the breather-soliton \( S_{s}^{(m)}(\theta _{12}) \) \( S \)-matrix
of the SG model \cite{ZZ}:
\begin{equation}
\label{bsm}
S^{(m)}_{s}(\theta )=\frac{\sinh \theta +i\cos \frac{\pi mp}{2}}{\sinh \theta -i\cos \frac{\pi mp}{2}}\prod ^{m-1}_{l=1}\frac{\sin ^{2}\left( \frac{m-2l}{2}\pi p-\frac{p}{4}+i\frac{\theta }{2}\right) }{\sin ^{2}\left( \frac{m-2l}{2}\pi p-\frac{p}{4}-i\frac{\theta }{2}\right) }\,.
\end{equation}
 In finite volume, when the boundary condition (\ref{quasi_periodic}) is imposed
the single kink particles disappear from the theory and they survive only as
the building block constituents of the multi-kink state with integer topological
charge. On the other hand the breather particles \( B_{n}^{(m)} \) are there
even in finite volume, as they are consistent with (\ref{quasi_periodic}).

\subsection{The Bethe--Yang equations \label{subsec:bethe_yang_equations}}

To make a comparison with the TCSA results we need the multi-particle energy
levels as functions of \( L \). While it is possible to get them directly
in the NLIE formalism, we shall use a simpler (and approximate) method which
can be shown to give equivalent results for large volume (\( ML\gg 1 \)) but
which has a clear interpretation in terms of the spectrum of the model as it
was described above.

In an integrable theory particle number is conserved, thus the concept of \( N \)
particle state with any fixed \( N \) is well defined, at least for \( L\gg M^{-1} \)
when most of the time the particles are far from each other. First we consider
the `pure' multi-kink states and envisage the \( N \)-kink energy eigenstate
in a large volume as a stationary scattering state, characterized by the set
of (conserved) rapidities \( \vec{\theta }=\{\theta _{1},\ldots ,\theta _{N}\} \).
Any \( N \)-kink stationary wavefunction can be expanded using the independent 
allowed \( N \)-kink \emph{in}-states (the number of which we
denote by \( d_{N} \))
\[
\left| K_{n_{1}n_{2}}(\theta _{1})K_{n_{2}n_{3}}(\theta _{2})\ldots
 K_{n_{N}n_{1}}(\theta _{N})\right\rangle \: ,\quad \theta _{1}>\theta
 _{2}>\ldots >\theta _{N}
\] 
as basis, and let \( \psi^{\mathbf{n}}(\vec{\theta }) \)
 ($\mathbf{n}=(n_{1}\ldots n_{N})$) denote its components with respect
 to this basis. Then, as a consequence of the periodic boundary
 conditions \( \psi ^{\mathbf{n}}(\vec{\theta }) \) must
 satisfy the following Bethe--Yang equations (at least if all particles
 have the same mass):
\begin{equation}
\label{BY}
e^{iML\sinh \theta _{j}}\sum _{\mathbf{n}}T_{j}( \vec{\theta }) ^{\mathbf{m}}_{\mathbf{n}}\psi ^{\mathbf{n}}( \vec{\theta }) =-\psi ^{\mathbf{m}}( \vec{\theta }) \, ,\quad j=1,\ldots ,N\, ,
\end{equation}
where \( T_{j}(\vec{\theta }) \) is the \( N \) particle transfer
matrix \cite{KM3}: 
\[
T_{j}( \vec{\theta })
^{\mathbf{m}}_{\mathbf{n}}=\prod
^{N}_{i=1}S_{m_{i}n_{i+1}}^{n_{i}m_{i+1}}\left( \theta _{j}-\theta
_{i}\right) \: .
\]
 For a given \( L \) eqns.~(\ref{BY}) have solutions only for some special
\( \theta _{j} \), and the total energy and momentum of the system in a state
characterized by these solutions are given by 
\begin{equation}
\label{BA_energy_momentum}
E=\sum ^{N}_{j=1}M\cosh \theta _{j}\: ,\quad P=\sum ^{N}_{j=1}M\sinh \theta _{j}\,.
\end{equation}
 Note however, that while this expression for \( P \) is exact even for finite
\( L \), the one for \( E \) is only approximate, as we neglect the `tunnelling'
and `off-shell' corrections.

It is straightforward to use this formalism to obtain the energy levels of pure
multi-kink states in the \( Q=n \) (\( n\geq 1 \)) sector of \( \mathrm{SG}(\beta ,k) \).
It is natural to assume, that the lowest energy levels in this sector correspond
to states with the smallest possible number of particles. The states with lowest
number of particles compatible with the boundary condition and the topological
charge being \( n \) contain \( N=k\cdot n \) kinks and no breathers or antikinks.
Since the sequence of kinks in these states is necessarily fixed, we have, independently
of \( n \), only \( k \) different basis vectors in this subspace, i.e.~\( d_{N}=k \).
The basis vectors can be chosen as 
\begin{eqnarray}
\left| \psi _{0,1,\ldots ,k-1}\left( \theta _{1},\ldots ,\theta _{k\cdot n}\right) \right\rangle  & = & \left| K_{01}\left( \theta _{1}\right) \ldots K_{k-1,0}\left( \theta _{k}\right) K_{01}\left( \theta _{k+1}\right) \ldots K_{k-1,0}\left( \theta _{k\cdot n}\right) \right\rangle \nonumber \\
\left| \psi _{k-1,0,\ldots ,k-2}\left( \theta _{1},\ldots ,\theta _{k\cdot n}\right) \right\rangle  & = & \left| K_{k-1,0}\left( \theta _{2}\right) K_{01}\left( \theta _{1}\right) \ldots \right. \nonumber \\
 &  & \left. K_{k-2,k-1}\left( \theta _{k}\right) K_{k-1,0}\left( \theta _{k+1}\right) \ldots K_{k-2,k-1}\left( \theta _{k\cdot n}\right) \right\rangle \nonumber \\
\left| \psi _{1,\ldots ,k-1,0}\left( \theta _{1},\ldots ,\theta _{k\cdot n}\right) \right\rangle  &=& \left| K_{1,2}\left( \theta _{1}\right) \ldots K_{0,1}\left( \theta _{k}\right) K_{1,2}\left( \theta _{k+1}\right) \ldots K_{0,1}\left( \theta _{k\cdot n}\right) \right\rangle\,, \label{KMexp} 
\end{eqnarray}
 where \( \theta _{1}>\theta _{2}>\ldots >\theta _{k\cdot n} \). Using the
explicit form of the transfer matrix and the identification of the various \( S_{\alpha \beta }^{\gamma \delta } \)
elements with the SG soliton-soliton \( S \)-matrix \( S_{ss}(\theta ) \),
the Bethe--Yang equations can finally be written in the following form:
\[
e^{iML\sinh \theta _{i}}\prod _{j\neq i}^{k\cdot n}e^{i\delta \left(
\theta _{i}-\theta _{j}\right) }(-1)^{k\cdot n}T_{k}\psi =-\psi \: ,\quad
i=1,\ldots ,k\cdot n\\,,
\]
where \( \psi  \) is a column vector made of the \( \psi^{\mathbf{n}} \)
coefficients of the basis vectors in eqn.~(\ref{KMexp}) and the \( k\times k \)
matrix \( T_{k} \)
\[
T_{k}=\left( \begin{array}{cccc}
0 & \ldots  &  & 1\\
1 & \ldots  &  & 0\\
0 & 1 & \ldots  & 0\\
\vdots  &  &  & \\
0 & \ldots  & 1 & 0
\end{array}\right) \]
 describes a cyclic permutation generating \( {\mathbb Z}_{k} \). The eigenvectors
\( \psi _{m} \) belonging to the various eigenvalues (\( e^{i\frac2\pi m/k} \),
\( m=0,\dots ,k-1 \)) of \( T_{k} \) carry the different inequivalent irreducible
representations of \( {\mathbb Z}_{k} \). Choosing the \( m \)-th eigenvalue
of \( T_{k} \) and introducing the dimensionless variable \( l=ML \) 
gives rise to the following Bethe--Yang equations 
\begin{equation}
\label{BAE}
l\sinh \theta _{i}+\sum _{j\neq i}\delta (\theta _{i}-\theta _{j})=2\pi \left( \hat{N}_{i}-\frac{m}{k}\right) \: ,i=1,\ldots k\cdot n\,,
\end{equation}
where \( \hat{N}_{i}\in {\mathbb Z}+1/2 \) for \( k\cdot n \) even
and \( \hat{N}_{i}\in {\mathbb Z} \) for \( k\cdot n \) odd, and as a result
of \( S_{ss}(0)=-1 \) we must have \( \hat{N}_{i}\neq \hat{N}_{j} \) for \( i\neq j \).
The total momentum carried by a solution of (\ref{BAE}) is determined solely
in terms of the \( \hat{N}_{i} \): 
\[ \frac{P}{M}=\frac{2\pi }{l}\left(\sum _{i}\hat{N}_{i}-m\cdot n\right)\,. \]
The simplest possibility---and the one we investigated by TCSA---is when the
total (CM) momentum vanishes, \( P=0 \). Please note that as a consequence
of \( \delta (-\theta )=-\delta (\theta ) \) and 
\(e^{i2\pi(k-m)/k}=e^{-i2\pi m/k} \)
the Bethe--Yang equations guarantee that the 
\( P=0 \) states \( (\psi _{m},\psi _{k-m}) \), which 
carry complex conjugate representations of \( {\mathbb D}_{k} \), are degenerate
in energy.

It is also possible to arrive at this result directly from the NLIE discussed
in Section \ref{sec:NLIE_description}. In the infrared limit \( ML\gg 1 \)
the integral term in the NLIE (\ref{alpha-nlie-excited}) becomes negligible and for
a state with \( k\cdot n \) holes the equation simplifies to
\[
Z(\theta )=\vartheta +l\sinh \theta +\sum ^{k\cdot n}_{j=1}\chi (\theta -\theta _{j})\: .\]
Observing that \( \chi (\theta )\equiv \delta (\theta ) \) and remembering
that the allowed values of \( \vartheta  \) (\ref{twist_values}) are exactly
equivalent to selecting
\[
\vartheta _{m}=2\pi \frac{m}{k}\: ,\quad k=0,\ldots ,m-1\: ,\] the
Bethe quantization rules reduce to eqn.~(\ref{BAE}) while the
energy/momentum formulas turn into (\ref{BA_energy_momentum}). Note
that the quantization rule (\ref{sG_NLIE_quant_rule}) which selects
the local operator algebra \( {\cal A}_{b} \) from
(\ref{local_algebras}) is exactly the one observed above for the \(
\hat{N}_{i} \): for \( k\cdot n \) even it assigns half-integer, while
for \( k\cdot n \) odd integer Bethe quantum numbers \( I_{i} \). We
remark here that in general the large volume limit of the NLIE
coincides with the Bethe--Yang equations (\ref{BY}) in their scalar
form, i.e.~evaluated on the eigenvectors of the transfer matrix \(
T^{\mathbf{m}}_{\mathbf{n}}( \vec{\theta }) \).

\subsubsection{Comparison with the TCSA data}

\begin{figure}
\centering
\includegraphics{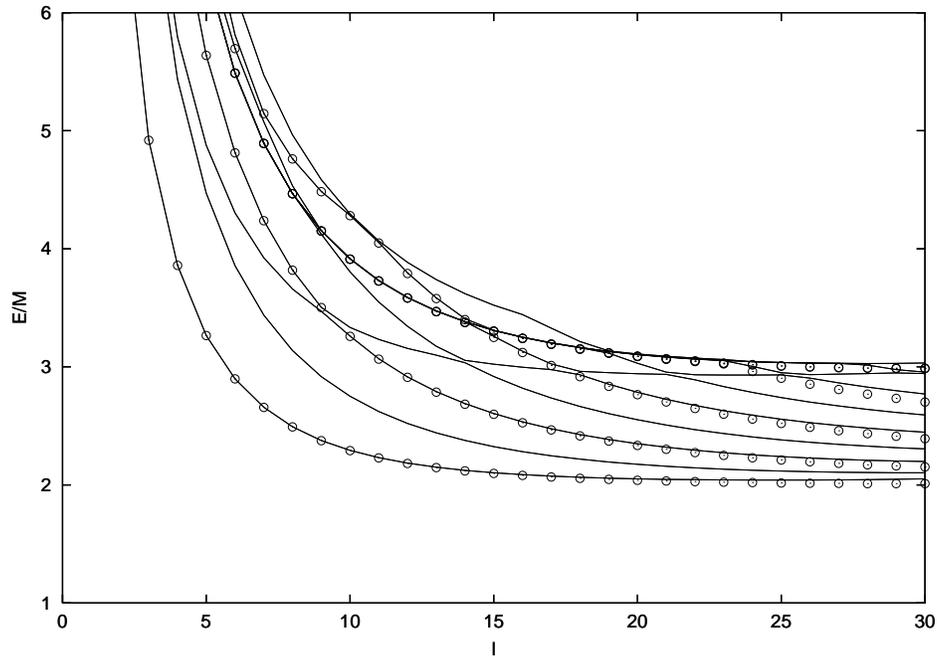}
\caption{Comparison of the spectra of
\(\mathrm{SG}\left(\beta=\frac{4}{3}\sqrt{\pi},2\right)_1\) 
and \(\mathrm{SG}\left(\beta=\frac{4}{3}\sqrt{\pi},1\right)_2\)}
\label{fig:missing}
\end{figure}

To compare with the TCSA results we consider a few sectors of the models with
\( k=2 \) and \( k=3 \) in more detail. The simplest of them is the \( Q=1 \)
sector of \( \mathrm{SG}(\beta ,2) \). In this case \( \psi _{0} \) describes the symmetric
and \( \psi _{1} \) the antisymmetric wave functions of \( K_{01} \) and \( K_{10} \),
the latter one being allowed as the two kinks are \emph{different} (bosonic)
particles. \( P=0 \) implies in this case that \( \theta _{1}=-\theta _{2}=\theta  \)
and (\ref{BAE}) simplify to 
\begin{equation}
\label{2eq}
l\sinh \theta +\delta (2\theta )=2\pi \left\{ \begin{array}{ll}
N_{0}\in {\mathbb Z}+\frac{1}{2}\, , & m=0\, ,\\
N_{1}\in {\mathbb Z} \setminus\{0\}\,, & m=1\, .
\end{array}\right. 
\end{equation}
 From this equation \( \theta =\theta (l,N_{m}) \) can be determined using
e.g.~an iterative procedure, or alternatively the volume dependence of the
\( 2 \)-kink energy levels can be given in parametric form as
\begin{equation}\label{2para}
\left( l,\frac{E}{M}\right) (\theta )=\left( \frac{2\pi N_{m}-\delta
(2\theta )}{\sinh \theta }\, ,\, 2\cosh \theta \right) \,
.\end{equation} 
This makes it clear that on the \( (l,E/M) \)
plane the \( 2 \)-particle lines cannot intersect each other for \(
0<l<\infty \).  These findings make it possible to distinguish clearly
between the solitons of sine--Gordon theory and the kinks in the
2-folded model, i.e.~to argue that the \( Q=2\) sector of \(
\mathrm{SG}(\beta ,1)\) (denoted as \( SG(\beta ,1)_2\)) is different
from the \( Q=1\) sector of \( \mathrm{SG}(\beta ,2)\) (\( SG(\beta
,2)_1\)). Indeed it is straightforward to derive the Bethe--Yang
equations for the pure 2-soliton states in \(\mathrm{SG}(\beta
,1)_2\); since these solitons are identical particles these Bethe--Yang
equations are given by the first line in eqn.~(\ref{2eq}). Therefore
the number of 2-soliton states in \(\mathrm{SG}(\beta
,1)_2\) is half the number of the 2-kink ones in \(\mathrm{SG}(\beta
,2)_1\). On Fig.~\ref{fig:missing} the continuous lines are given by
the \textit{interpolated} TCSA data obtained in
\(\mathrm{SG}(\beta,2)_1\) with $p=2/7$, while the dots represent the
TCSA data obtained in \(\mathrm{SG}(\beta ,1)_2\); they clearly
correspond to every second line only\footnote{This is also true
for the two lines which cannot be interpreted as two-particle ones due
to their (multiple) intersections with the other levels.}.
\begin{figure}
\centering
\includegraphics{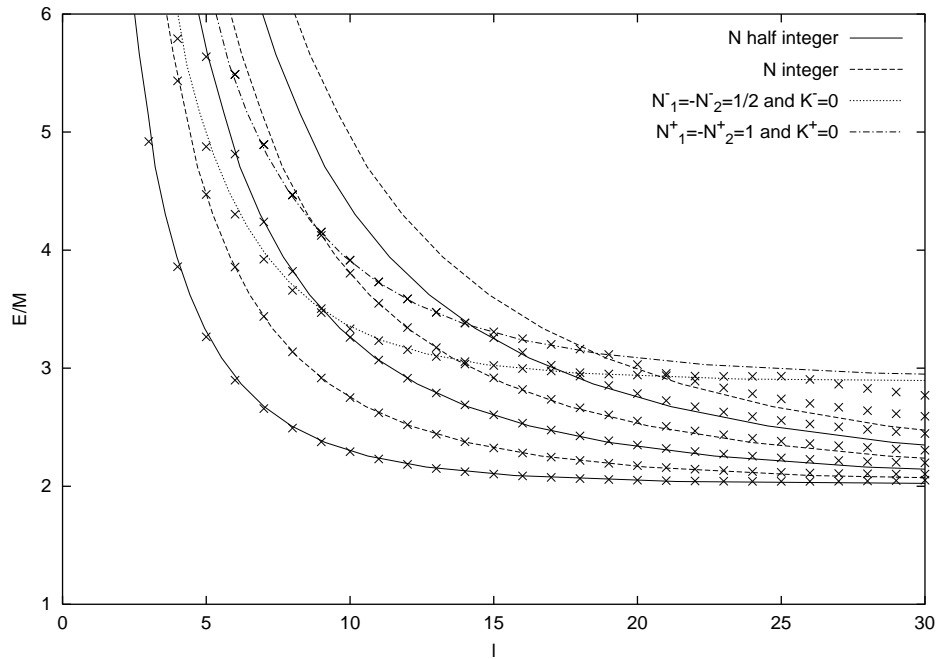}
\caption{The full spectrum and the predictions for \(\mathrm{SG}\left(\beta=\frac{4}{3}\sqrt{\pi},2\right)_1\)}
\label{fig:KM2}
\end{figure}

On Fig.~\ref{fig:KM2} the continuous lines depending on one quantum
number only are given by eqn.~(\ref{2eq}) and the dots now correspond
to the TCSA data in \(\mathrm{SG}(\beta ,2)_1\). On this figure we find data
lines that cannot be interpreted as pure \( 2 \)-kink states, as they
apparently do intersect some of the other lines. Since the large \( l
\) behaviour of these lines is compatible with \( E\rightarrow
2M+M_{1} \) for \( l\rightarrow \infty \), it is natural to try to
interpret them as describing \( 3 \)-particle states containing one
`first' (\( m=1 \)) breather in addition to the two kinks. Since \(
M_{1}\neq M \) the transfer matrix formalism worked out in \cite{KM3}
does not apply directly. Nevertheless following the original line of
thought, namely by `commuting around' any of the particles using the
appropriate \( S \)-matrices one can derive the Bethe--Yang equations
for the stationary scattering states.

Deleting the sequential or `upper' \( (m) \) index of the breathers (since
we consider only the first one) but keeping their `vacuum' index the basis vectors
in this \( 3 \)-particle subspace can be chosen as 

\begin{eqnarray*}
\psi ^{01\mathrm{h}} & = & \left| K_{01}\left( \theta _{1}\right) K_{10}\left( \theta _{2}\right) B_{0}\left( \theta _{3}\right) \right\rangle \\
\psi ^{10\mathrm{h}} & = & \left| K_{10}\left( \theta _{1}\right)
K_{01}\left( \theta _{2}\right) B_{1}\left( \theta _{3}\right)
\right\rangle\,,\quad \theta _{1}>\theta _{2}>\theta _{3}\: ,
\end{eqnarray*}
 plus \( 4 \) other states corresponding to the breathers being `in the middle'
or in the `front'. (The index \( \mathrm{h} \) signals that the breathers are
the last). Introducing the notation \( S_{s}^{(1)}(\theta )=-e^{i\alpha (\theta )} \),
\( M_{1}L=ml \) and \( \psi ^{\pm }=\frac{1}{\sqrt{2}}(\psi ^{01\mathrm{h}}\pm \psi ^{10\mathrm{h}}) \),
one can convert the Bethe--Yang equations derived by the `commuting around' procedure
into the following form: 
\begin{eqnarray}
l\sinh \theta _{1}+\delta (\theta _{1}-\theta _{2})+\alpha (\theta _{1}-\theta _{3}) & = & 2\pi N_{1}^{\gamma }\label{BYbdiag1} \\
l\sinh \theta _{2}+\delta (\theta _{2}-\theta _{1})+\alpha (\theta _{2}-\theta _{3}) & = & 2\pi N_{2}^{\gamma }\label{BYbdiag2} \\
ml\sinh \theta _{3}+\alpha (\theta _{3}-\theta _{1})+\alpha (\theta _{3}-\theta _{2}) & = & 2\pi K^{\gamma },\label{BYbdiag3} 
\end{eqnarray}
 where \( \gamma =\pm  \), \( N_{1,2}^{+}\in {\mathbb Z} \), \( N_{1,2}^{-}\in {\mathbb Z}+1/2 \),
\( K^{\gamma }\in {\mathbb Z} \). Once again, an equivalent system of equations
can be derived from the NLIE in the \( ML\gg 1 \) (infrared) region.

In the \( P=0 \) system there are solutions with the breather at rest and the
two kinks moving in opposite directions: \( K^{\gamma }=0=\theta _{3} \), \( N_{2}^{\gamma }=-N_{1}^{\gamma } \)
\( \theta _{2}=-\theta _{1} \). Then the (\ref{BYbdiag1}--\ref{BYbdiag3})
system simplifies to the single equation 
\begin{equation}
\label{kkbr}
l\sinh \theta _{1}+\delta (2\theta _{1})+\alpha (\theta _{1})=2\pi N_{1}^{\gamma },
\end{equation}
 which admits the parametric solution 
\begin{equation}
\label{eqn:param2b}
\left( l,\frac{E}{M}\right) (\theta )=\left( \frac{2\pi N_{1}^{\gamma }-\delta (2\theta _{1})-\alpha (\theta _{1})}{\sinh \theta _{1}},2\cosh \theta _{1}+m\right) .
\end{equation}
 It is easy to understand how the \( N_{1}^{-}=-N_{2}^{-}=1/2 \), \( K^{-}=0 \)
state tends to the \( N_{1}=1 \) \( 2 \)-kink state (eqn.~(\ref{2eq})) in
the UV limit: in this limit \( \alpha (\theta )\rightarrow -\pi  \) and this
effectively converts the \( 1/2 \) on the right hand side of (\ref{kkbr})
into \( 1 \). The lines depending on three quantum numbers on
Fig.~\ref{fig:KM2} correspond to
the first two possibilities given by 
eqn.~(\ref{kkbr}), and the agreement with the TCSA data is
excellent. Note also that the line with $N_1^+=N_2^+=1$, $K^+=0$ is
also present in $\mathrm{SG}(\beta,1)_2$, consistently with the data
on Fig.~\ref{fig:missing}.     

\begin{figure}
\centering
\includegraphics{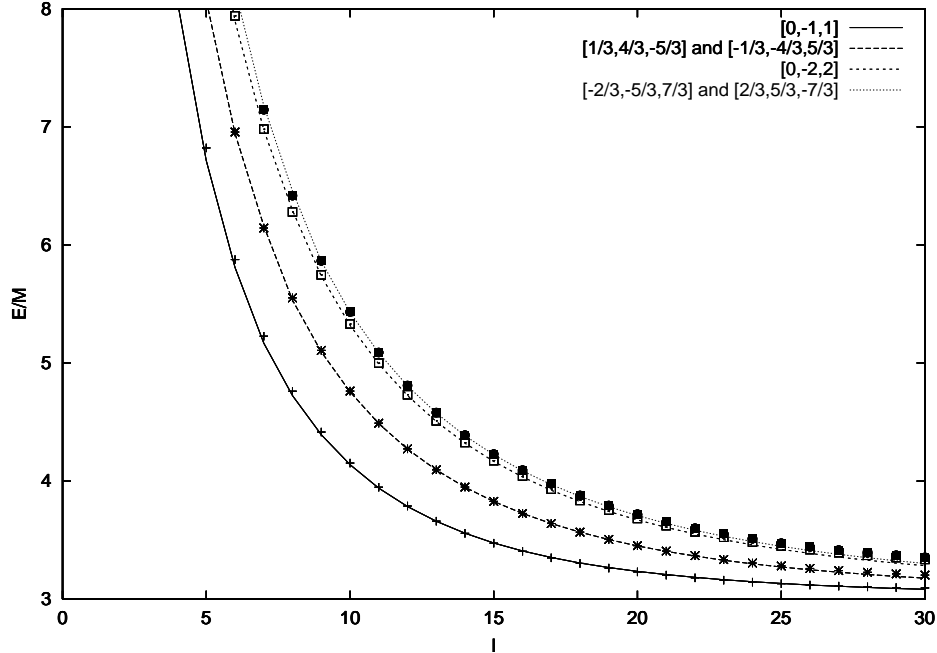}
\caption{Bethe--Yang curves and TCSA data for \(\mathrm{SG}\left(\beta=\frac{4}{3}\sqrt{\pi},3\right)_1\)}
\label{fig:KM3}
\end{figure}

The \( Q=1 \) sector of \( \mathrm{SG}(\beta ,3) \) is interesting, as
the representations belonging to \( m=1 \) and \( m=2\equiv -1 \) are complex
conjugate ones. The Bethe--Yang equations in this case take the form 
\[
l\sinh \theta _{i}+\sum _{j\neq i}^{3}\delta \left( \theta _{i}-\theta _{j}\right) =2\pi \left\{ \begin{array}{ll}
N^{(i)}_{0}\in {\mathbb Z}\, , & m=0\, ,\\
N^{(i)}_{\pm 1}\in {\mathbb Z}\pm \frac{1}{3}\, , & m=\pm 1\, ,
\end{array}\right. \:\mathrm{where}\:
\begin{array}{l} i=1,2,3 \ \mathrm{and}\\ 
N_{m}^{(i)}\neq N_{m}^{(j)}\mathrm\:\mathrm{for}\: i\neq j
\end{array}\, .
\]
For the symmetric states (i.e.~when \( m=0 \)) with \( P=0 \) there
are special solutions with \( \theta _{3}=-\theta _{1} \), \( \theta
_{2}=0 \) (when \( N^{3}_{0}=-N^{1}_{0}=-N \), \( N^{2}_{0}=0 \)); and
for them the volume dependence of the energy levels can be given in
parametric form similar to eqn.~(\ref{2para}), but apart from these
special cases we have to rely on numerical procedures to get \(
\frac{E}{M}(l,N^{i}_{m}) \). On Fig.~\ref{fig:KM3}, for $p=2/7$, we
show how well the predictions of these Bethe--Yang equations describe
the TCSA data, regarding both the degeneracies and the volume
dependence.

\section{Vacuum expectation values of local fields}

In this section we analyze the vacuum expectation values of
exponential fields. Some time ago an explicit expression was given for
this quantity by Lukyanov and Zamolodchikov \cite{lz_vevs}. Here we
compare this prediction with the data extracted from TCSA, and by
doing so we give further evidence that everything which in
sine--Gordon theory follows only from the local properties of the
scalar field and the Lagrangian remains true in the \( k\)-folded
model as well.

\subsection{Symmetries and vacuum expectation values}

In this subsection we will isolate the independent amplitudes which characterize
the vacuum expectation values of exponential fields in \( \mathrm{SG}(\beta
,k)\):
\[
\left\langle \vartheta _{n}\right| V_{m}\left| \vartheta
_{r}\right\rangle \,,
\quad V_{m}=V_{(m,0)}=\exp \left( {i\beta }\frac{m}{k}\varphi \right) \: .\]
Our considerations will be valid for any value of the volume \( L \). 

As a consequence of the action of \( T \) on the vacua \( \left| \vartheta _{n}\right\rangle  \)
(\ref{T_theta_vacua_action}) and on the local fields \( V_{m} \) (\ref{ST_action_on_vertex_ops})
we have 
\begin{equation}
\label{theta_vevs}
\left\langle \vartheta _{n}\right| V_{m}\left| \vartheta _{r}\right\rangle =A_{m}(r)\delta _{n,m+r}\,,\quad \left\langle n\right| V_{m}\left| r\right\rangle =B_{m}(r-n)e^{\frac{2\pi i}{k}mn}\,,
\end{equation}
 where \( A_{m}(r) \) are certain (unknown) amplitudes and 
\begin{equation}
\label{n_vevs}
B_{m}(r)=\left\langle 0\right| V_{m}\left| r\right\rangle =\frac{1}{k}\sum ^{k-1}_{n=0}e^{-\frac{2\pi i}{k}nr}A_{m}(n)\,.
\end{equation}
Reality of the field \( \varphi  \) yields 
\[
V^{\dagger }_{m}=V_{-m}\]
and so
\[
B_{m}(r)^{*}=e^{\frac{2\pi i}{k}rm}B_{-m}(-r)\,.\]
 The transformation properties of the field $\varphi$ under \( S \) imply
\[
B_{m}(r)=B_{-m}(-r)\,.\]
This allows us to determine the phase of the amplitude \( B_{m}(r) \) up to
a sign:
\begin{equation}
\label{vev_phase}
B_{m}(r)=e^{\frac{\pi i}{k}rm}F_{m}(r)\,,\quad F_{m}(r)\in {\mathbb R}\,.
\end{equation}
It turns out that the real amplitudes \( F_{m}(r) \) are not all
independent.  Indeed, as it was already remarked in subsection
\ref{subsec:tcsa_description}, in the usual basis (\ref{Fock_space})
of the (ultraviolet) free boson Hilbert space all the \( V_{m} \) have
real matrix elements. Therefore the Hamiltonian (\ref{sG_Hamiltonian})
as a matrix is real and symmetric, and as a consequence all its
eigenvectors have real components.  This implies
\[
\left\langle \vartheta _{n}\right| V_{m}\left| \vartheta _{r}\right\rangle \in {\mathbb R}\]
and as a result
\begin{equation}
\label{vev_reality}
F_{m}(r)=(-1)^{m}F_{m}(k-r)\,,\quad r=1,\ldots ,k-1\,.
\end{equation}
Note that there is no constraint on \( F_{m}(0)=\left\langle 0\right| V_{m}\left| 0\right\rangle  \).
The vacuum expectation values can therefore be characterized by the independent
amplitudes
\begin{equation}
\label{vev_independent}
F_{m}(r)\,,\quad r=0,\ldots ,\left[ \frac{k}{2}\right] 
\end{equation}
all of which are real.

\subsection{The Lukyanov--Zamolodchikov formula}

In \cite{lz_vevs} Lukyanov and Zamolodchikov proposed 
an exact formula for the vacuum expectation values of exponential
fields in the ordinary sine--Gordon model in infinite volume. 
Here we briefly recall their result. We normalize
the exponential fields so that the short distance asymptotics of the non-vanishing
two-point functions reads
\[
\left\langle 0\right| e^{ia\varphi (x)}e^{-ia\varphi (y)}\left|
0\right\rangle =\frac{1}{|x-y|^{2\Delta _{a}}}\,, \]
where \( \Delta _{a}=\frac{a^{2}}{8\pi } \) is the conformal weight of
\( e^{ia\varphi (x)} \), and $\left| 0\right\rangle$ denotes the state
$\left| n\right\rangle$ for $n=0$ (see eqn.~(\ref{n_vacua})). Let us define 
\[
{\cal G}(a)=\left\langle 0\right| e^{ia\varphi (x)}\left|
0\right\rangle\,.
\]
The authors of \cite{lz_vevs} conjecture\footnote{Our normalization
for the field \( \varphi \) and the coupling constant \( \beta \)
differs from that of \cite{lz_vevs} by a factor of \( \sqrt{8\pi } \).
}

\begin{eqnarray}
{\cal G}(a)& 
= & \left[ \frac{M\sqrt{\pi }\Gamma \left( \frac{p+1}{2}\right) 
 }{2\Gamma \left(\frac{p}{2}\right)}\right] ^{\frac{a^{2}}{4\pi
}}\times \nonumber \\
 &  & \exp \left\{ \int ^{\infty }_{0}\frac{\ud t}{t}\left[ \frac{\sinh ^{2}\left( \frac{a\beta }{4\pi }t\right) }{2\sinh \left( \frac{p}{p+1}t\right) \sinh (t)\cosh \left( \frac{1}{p+1}t\right) }-\frac{a^{2}}{4\pi }e^{-2t}\right] \right\}\,, 
\label{exactvev}\end{eqnarray}
which is valid for 
\begin{equation}\label{condts} 
\beta ^{2}<8\pi  \quad {\mathrm{and}} \quad \left| \Re e\, a\beta
\right| <4\pi\,,  
\end{equation}
and where $M$ is the soliton (kink) mass.  Recalling now our basic
idea, namely that everything which depends only on the local
properties of $\varphi $ and the SG Lagrangian also holds for the
$k$-folded model $\mathrm{SG}(\beta ,k)$, we can identify \({\cal
G}(a)\) and \( F_m(0)\):
\[
{\cal G}\left( \frac{m\beta }{k}\right) =F_{m}(0)\quad
\mathrm{when}\quad l=ML=\infty .\]
The TCSA data allow us to extract at finite \( l \)  the
expectation value of \( e^{ia\varphi (w,\bar{w})}\vert_{\vert w\vert
=1}\) (where \( w=e^{\frac{2\pi}{L}z}\)  is the map from the cylinder
to the plane) in the ground states found by the numerical
diagonalization, i.e.~we can measure \( A_m(n)\) as functions of \(
l\), \( A_m(n)[l]\). Thus introducing the dimensionless function
\(g(a)=M^{-2\Delta_a}{\cal G}(a)\) we obtain for finite \( l\):  
\begin{equation}\label{ourF}
F_m(0)[l]=\frac{1}{k}\sum\limits_{n=0}^{k-1}A_m(n)[l]=\frac{l^{2\Delta_a}}{(2\pi
)^{2\Delta_a}}g(a)N(l)\,,\quad a=\frac{m\beta}{k}\,,
\end{equation}
where \( N(l)\) is a finite size ``correction'' factor of which we know \(
N(l)\rightarrow 1+{\cal O} (e^{-l})\) for \(l\rightarrow\infty\).

The amplitudes \( F_{m}(r) \) are in general not known for \( r\neq 0 \).
However, one can derive their leading behaviour for large volume using the following
consideration. First note that the matrix elements 
\[
\left\langle 0\right| V_{m}\left| r\right\rangle \]
vanish in infinite volume since there is a super-selection rule making the vacua
\( \left| n\right\rangle  \) lie in physically disconnected Hilbert spaces
with no local operators connecting them. It is clear that in finite volume the
non-vanishing contribution to these matrix elements comes from the vacuum tunnelling
described by instanton effects, whose magnitude we calculated for large \( l \)
in section \ref{sec:instantons}. Therefore we expect 
\[
\left\langle 0\right| V_{m}\left| r\right\rangle \sim e^{-l}\quad \mathrm{when}\quad l\gg 1\quad \mathrm{and}\quad r\neq 0\, .\]

\subsection{Comparison to TCSA}

\begin{figure}
\centering \includegraphics{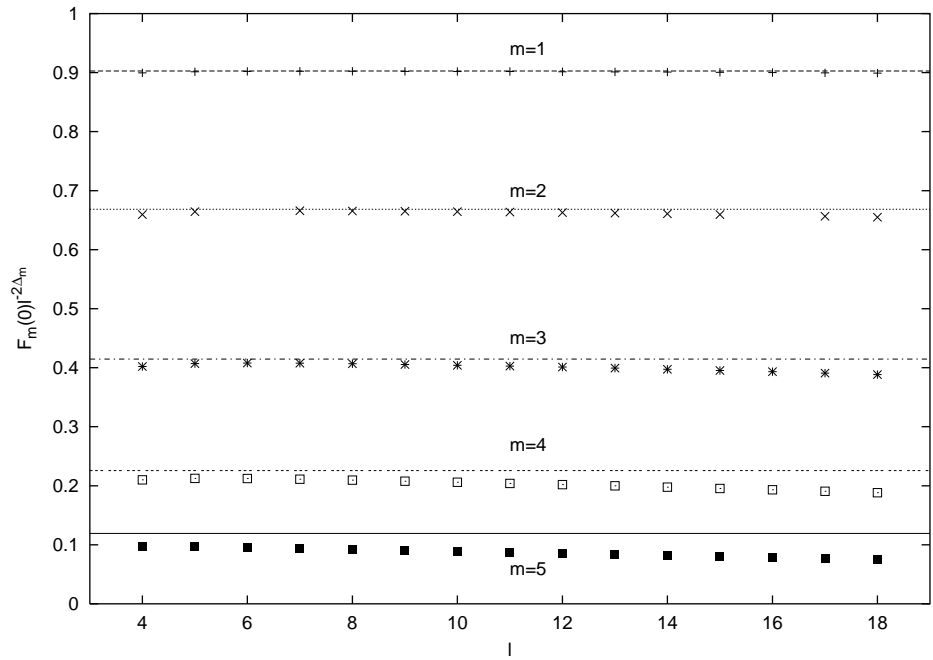}
\caption{TCSA data and the Lukyanov--Zamolodchikov prediction}\label{vacexp}
\end{figure}

We tested the Lukyanov--Zamolodchikov formula in a 3-folded model with
$p=2/7$. Using the eigenvectors belonging to the three ground states
found by the TCSA algorithm we determined numerically the various
amplitudes $A_m(n)[l]$ for all the values of $m$ ($m=1,\dots ,5$)
satisfying the conditions (\ref{condts}). Then, recalling
eqn.~(\ref{ourF}), we plotted $F_m(0)[l]l^{-2\Delta_m}$ where
\[ 
2\Delta_m=\frac{m^2\beta^2}{4k^2\pi}
\]
as a function of $l$. The result is shown on Fig.~\ref{vacexp}, where the
horizontal lines correspond to 
\[ 
(2\pi)^{-2\Delta_m}g\left(\frac{m\beta}{k}\right)\,.
\]
The agreement between the measured values and the
predicted ones is very good, though for higher values of $m$ the data
are somewhat below the horizontal lines indicating that the TCSA data
should be extrapolated as in the discussion that follows.
\begin{figure}
\centering \includegraphics{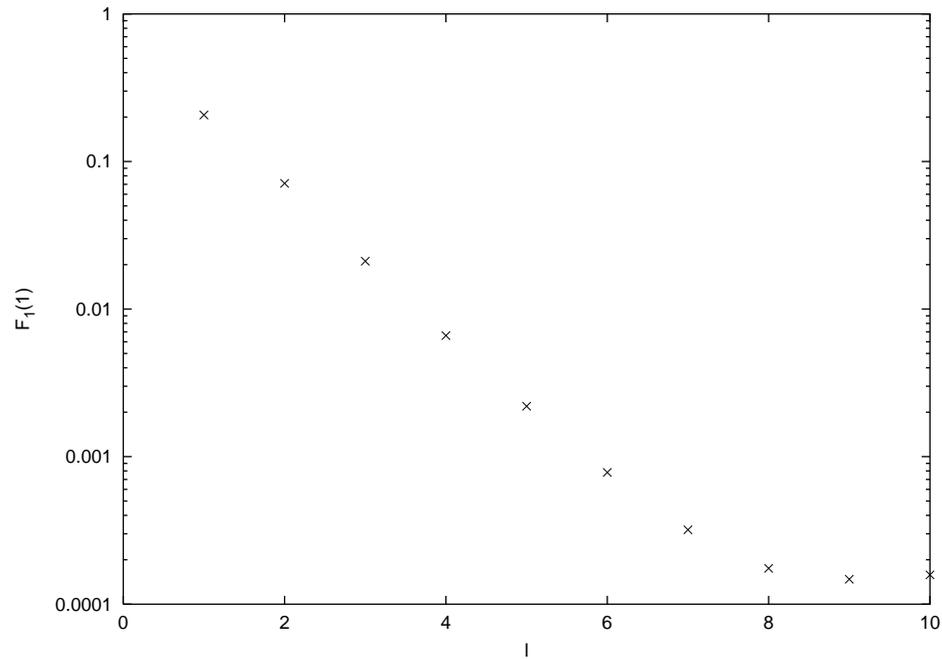}
\caption{The decay of $F_1(1)[l]$ }\label{decay}
\end{figure}
\begin{figure}
\psfrag{VEV}{$\mathrm F_1(0)/\mathrm l^{2\Delta_1}$}
\centering \includegraphics{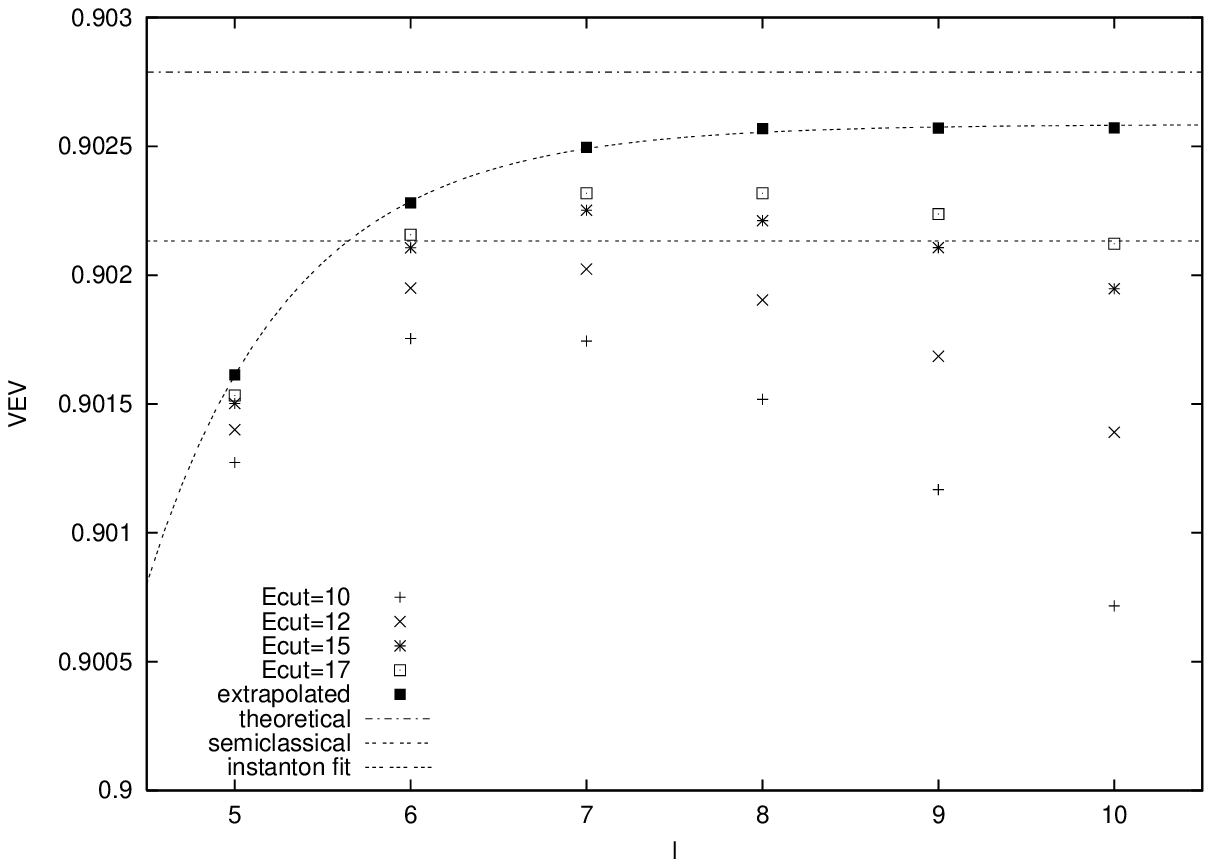}
\caption{Comparing TCSA with the semiclassical and full quantum
formulae}
\label{VEVfit}
\end{figure}
The 3-folded model provides a good laboratory as it makes possible to
extract $F_m(1)[l]$ from the data as well. The result 
$F_1(1)[l]$ can be seen on a semilogarithmic plot on
Fig.~\ref{decay}. Clearly this behaviour is consistent with the
expected exponential fall-off. 

The exact expression for \({\cal G}(a)\) was obtained in
\cite{lz_vevs} by a clever interpolation between various limits, where
the corresponding expressions were known from other sources. One
important such limit was the semiclassical one. Since our numerical
study is for \( p<1\), we investigated whether using our data one can
make a distinction between the exact and semiclassical expressions of
\cite{lz_vevs}, i.e.~whether one can justify the exact or merely the
semiclassical formula. 

We carried this out by zooming in on the
vicinity of the uppermost (\( m=1\)) line on Fig.~\ref{vacexp}, and
the results are compiled on Fig.~\ref{VEVfit}. Here the upper/lower
horizontal line corresponds to a \( g(a,\beta)\) obtained by the
exact/semiclassical expressions in  \cite{lz_vevs}, and the TCSA data
displayed were taken at four different \( E_{\mathrm{cut}}\) values. At
each fixed $l$ the data apparently converge monotonously with
increasing \(E_{\mathrm{cut}}\). The validity of the exact expression over
the semiclassical one is supported by the fact that these monotonously
convergent data exceed the semiclassical expression in a certain
range, while they always stay below the exact one. To strengthen this
conclusion, at each fixed $l$, we extrapolated the data by fitting
their \(E_{\mathrm{cut}}\) dependence with an expression
\[
a(l)\frac{E_{\mathrm{cut}}^{2(2-h)}+c(l)}{E_{\mathrm{cut}}^{2(2-h)}+b(l)}\,,
\] 
and the extrapolated points on
Fig.~\ref{VEVfit} correspond to the coefficients \( a(l)\).\footnote{This
extrapolating formula is consistent with the monotonous increase and
the power of \(E_{\mathrm{cut}}\) in it is motivated by the
observation, that envisaging the determination of \({\cal G}(a)\) in
pCFT the effective expansion parameter would be \(\lambda^2\) rather
than \(\lambda\).}  These extrapolated points show a monotonously
increasing behaviour in $l$ and they significantly exceed the
semiclassical expression while always stay below the exact one. In the
final step, recalling eqn.~(\ref{ourF}), we fitted the extrapolated
points by \( A-Be^{-l}/l\), where \( e^{-l}\) is motivated by
the instanton contribution, and the \( l^{-1}\) by the fluctuation
determinant \textit{without} a zero mode. This \lq instanton fit'
describes the extrapolated data with a very small variance, and though
the $A$ obtained this way is somewhat smaller than the exact
expression, it is much closer to the exact than to the semiclassical
one. To sum up, we can say that our data indeed favour the exact
expression of \cite{lz_vevs} over the semiclassical one.

We remark that a similar calculation of vacuum expectation values from
TCSA was performed in \cite{tcs_vevs} for the case of $\Phi_{(1,3)}$
perturbations of Virasoro minimal models. These models are
restrictions of sine--Gordon theory at rational values of $p$ and as a
consequence one can derive a formula for the vacuum expectation values
of local fields starting from (\ref{exactvev})
\cite{lz_vevs}. Similarly to our case, they
find that the respective formula agrees with the TCSA results. In
contrast to their approach, however, we checked the prediction 
(\ref{exactvev}) directly in the case of sine--Gordon theory.

There is another important implication of this result. The
Lukyanov--Zamolodchikov conjecture for the vacuum expectation values of
local fields is connected to (and can be derived from) the conjecture
for the so-called Liouville reflection factor \cite{l3pt, vevs_refl}. The
above verification of the formula (\ref{exactvev}) therefore lends an
indirect support to the conjectured expression for the reflection
factor and considerations based on it.

\section{Conclusions}

In this paper we investigated the $k$-folded sine--Gordon model
$\mathrm{SG}(\beta ,k)$ in finite volume. The aim of this study is to give
support to the idea that the $k$-folded boundary conditions which make    
$\mathrm{SG}(\beta ,k)$ different from the ordinary sine--Gordon theory do indeed
preserve the integrability of the model, while changing the spectrum
in a well defined manner.

We analyzed three major problems in some detail and showed in all of
them, that the consequences one can draw from this expectation are
indeed correct. In particular we found that the leading part of the
split in the ground state energy levels, for which the NLIE and the
instanton calculus gave identical results, does indeed coincide with
the numerical data obtained by using TCSA. Furthermore, we provided
evidence that the $k$ dependent degeneracies and the volume
dependence of the multi-particle energy levels can indeed be described
by the formalism of \cite{KM3}, thus indirectly we verified the
conjectured $S$-matrices. Last but not least we showed that the vacuum
expectation value of the exponential field can be measured in the $k$
folded model and we gave evidence supporting the validity of the formula 
proposed in \cite{lz_vevs}. 
\vspace{1cm}

\begin{center}\textbf{Acknowledgements}\end{center}
G. Tak\'acs would like to thank G. Watts for useful
discussions. G. T. is supported by a PPARC (UK) postdoctoral
fellowship, while Z. B. by an OTKA (Hungary) postdoctoral fellowship
D25517/97. This research was also supported in part by the Hungarian
Ministry of Education under FKFP 0178/1999 and the Hungarian National
Science Fund (OTKA) T029802/99.

\vspace{1cm}
\appendix\renewcommand\theequation{\hbox{\normalsize\Alph{section}.\arabic{equation}}}

\section{Instanton calculus in finite volume \label{app:instanton_calculus}}

In this appendix we perform a standard instanton calculation based on the dilute
instanton gas approximation, following the outlines of \cite{munster}.
Here we shall work in a \( k=\infty  \) theory which means that we drop the
identification of the field in (\ref{period}). We will comment on this issue
later.

\subsection{The dilute instanton gas approximation}

The Euclidean action of sine--Gordon theory is
\begin{equation}
\label{euclidean_action}
S_{E}[\varphi ]=\int ^{\infty }_{-\infty }\ud\tau \int ^{L/2}_{-L/2}\ud x\left( \frac{1}{2}\left( \partial _{\tau }\varphi \right) ^{2}+\frac{1}{2}\left( \partial _{x}\varphi \right) ^{2}+\frac{\mu ^{2}_{0}}{\beta ^{2}}\left( 1-\cos \beta \varphi \right) \right)\,. 
\end{equation}
The equations of motion following from the action (\ref{euclidean_action})
admit the one-instanton solution
\begin{equation}
\label{instanton_solution}
\varphi _{\mathrm{inst}}=\frac{4}{\beta }\arctan \exp \left( \mu _{0}\left( \tau -\tau _{0}\right) \right) 
\end{equation}
with action
\begin{equation}
S_{E}\left[ \varphi _{\mathrm{inst}}\right] =\frac{8\mu _{0}}{\beta ^{2}}L=M_{\mathrm{class}}L\,,
\end{equation}
where \( M_{\mathrm{class}} \) is the classical soliton mass. Using
the usual rules of instanton calculus, the saddle-point evaluation of
the Euclidean path integral yields the following result for the level
splitting as a function of the twist angle \( \vartheta \) labelling
the energy eigenstates $|\vartheta\rangle$:
\begin{equation}
\label{saddle_point_formula}
\Delta E(\vartheta )=-2\cos \left( \vartheta \right) \left| \frac{\det 'M}{\det M_{0}}\right| ^{-1/2}\left( \frac{S_{E}\left[ \varphi _{\mathrm{inst}}\right] }{2\pi }\right) ^{1/2}e^{-S_{E}\left[ \varphi _{\mathrm{inst}}\right] }\,,
\end{equation}
where \( M \) and \( M_{0} \) are the operators 
\begin{equation}
M=-\partial ^{2}_{\tau }-\partial ^{2}_{x}+V''\left( \varphi
_{\mathrm{inst}}\right) 
\,,\quad M_{0}=-\partial ^{2}_{\tau }-\partial ^{2}_{x}+V''\left( 0\right) 
\end{equation}
describing the fluctuations around the instanton to quadratic order, \( \det 'M \)
denotes the determinant of \( M \) without its zero mode and the factor 
\begin{equation}
\label{zeromode_part}
\left( \frac{S_{E}\left[ \varphi _{\mathrm{inst}}\right] }{2\pi }\right) ^{1/2}
\end{equation}
comes from the one translational zero mode \( \tau _{0} \) of the instanton
solution (\ref{instanton_solution}). The \( \cos \vartheta  \) dependence
arises from Fourier transforming the dilute instanton gas summation according
to the relation between the vacua \( \left| n\right\rangle  \) and \( \left| \vartheta \right\rangle  \):
\begin{equation}
\left| \vartheta \right\rangle =\sum ^{\infty }_{n=-\infty }e^{in\vartheta }\left| n\right\rangle \: .
\end{equation}
 Here \( \vartheta  \) can take any value: the physically inequivalent choices
are \( -\pi <\vartheta \leq \pi  \). For \( \mathrm{SG}(\beta ,k) \), the result of
the dilute instanton gas calculation is similar to (\ref{saddle_point_formula})
with the only difference that \( \vartheta  \) can only take the discrete values
\[
\vartheta _{n}=\frac{2\pi n}{k}\: \bmod \: 2\pi \, .\]

\subsection{The heat kernel representation for the determinant}

Let us define the heat kernel for a positive hermitian operator \( A \) in
the following way:
\begin{equation}
K_{t}(A)=\mathrm{Tr}\, e^{-tA}\, ,
\end{equation}
from which the determinant can be reconstructed:
\begin{equation}
\log \det \mathrm{A}=\mathrm{Tr}\log A=-\int ^{\infty }_{0}\frac{\ud t}{t}K_{t}(A)\, .
\end{equation}
 As the determinant is divergent (the divergence comes from the lower end of
the integration over \( t \)), we shall compute the difference
\[
\tilde{K}_{t}(M)=K_{t}(M)-K_{t}\left( M_{0}\right) \]
using \( \zeta  \)-function regularisation. We define the \( \zeta  \)-function
\begin{equation}
\label{zeta_regularization}
\zeta (z,M)=\frac{1}{\Gamma (z)}\int ^{\infty }_{0}\ud t\ t^{z-1}\left( \tilde{K}_{t}(M)-1\right) \: .
\end{equation}
Then the determinant can be expressed as
\begin{equation}
\left| \frac{\det 'M}{\det M_{0}}\right| ^{-1/2}=\left. \exp \left( \frac{1}{2}\frac{d}{dz}\zeta (z,M)\right) \right| _{z=0}\: .
\end{equation}
We can separate the \( x \) and \( \tau  \) dependence and rewrite the heat
kernel in the following form:
\begin{eqnarray*}
\tilde{K}_{t}(M) & = & K_{t}\left( -\partial ^{2}_{x}\right) \left( K_{t}\left( Q\right) -K_{t}\left( Q_{0}\right) \right)\,, \\
Q & = & -\partial ^{2}_{\tau }+\mu ^{2}_{0}\left( 1-\frac{2}{\cosh ^{2}\left( \mu _{0}(\tau -\tau _{0})\right) }\right)\,, \\
Q_{0} & = & -\partial ^{2}_{\tau }+\mu ^{2}_{0}\,.
\end{eqnarray*}
The heat kernels for the three operators that appear above are
\begin{eqnarray}
K_{t}\left( -\partial ^{2}_{x}\right)  & = & \sum ^{\infty }_{n=-\infty }e^{-\left( \frac{2\pi }{L}n\right) ^{2}t}=\frac{L}{\sqrt{4\pi t}}\sum ^{\infty }_{n=-\infty }e^{-\frac{n^{2}L^{2}}{4t}}\,,\nonumber \\
K_{t}(Q) & = & 1+\int ^{\infty }_{-\infty }\ud p\rho (p)e^{-t\left( \mu ^{2}_{0}+p^{2}\right) }\,,\nonumber \\
K_{t}\left( Q_{0}\right)  & = & \int ^{\infty }_{-\infty }\ud p\rho _{0}(p)e^{-t\left( \mu ^{2}_{0}+p^{2}\right) }\,.\label{heat_kernel} 
\end{eqnarray}
where \( \rho  \), \( \rho _{0} \) are the spectral densities for the operators
\( Q \), \( Q_{0} \) and the additive \( 1 \) in \( K_{t}(Q) \) comes from
the zero mode.

\subsection{Evaluation of the spectral densities}

The spectral density of the operator 
\begin{equation}
Q=-\partial ^{2}_{\tau }+\mu ^{2}_{0}\left( 1-\frac{2}{\cosh ^{2}\left( \mu _{0}\tau \right) }\right) 
\end{equation}
 can be evaluated by solving the spectral problem:
\begin{equation}
\label{spectral_problem}
-\Psi ''-\frac{s(s+1)\mu ^{2}_{0}}{\cosh ^{2}\left( \mu _{0}\tau \right) }\Psi +\mu ^{2}_{0}\Psi =\lambda \Psi \: ,
\end{equation}
 where for our case \( s=1 \). This problem can be solved exactly by mapping the
above equation to a hypergeometric one \cite{landau}.

\subsubsection{The discrete spectrum}

The discrete spectrum of (\ref{spectral_problem}) corresponds to \( 0\leq \lambda <\mu ^{2}_{0} \).
The condition for square integrability reads
\begin{equation}
\label{discrete_spectrum_condition}
\sqrt{1-\frac{\lambda }{\mu ^{2}_{0}}}-s=-n\,,\quad n\in {\mathbb N}\: .
\end{equation}
For our case (\( s=1 \)) the only solution of (\ref{discrete_spectrum_condition})
is \( n=0 \) which corresponds to \( \lambda =0 \) i.e.~exactly the unique
zero mode of the operator \( Q \) mentioned before. (The other possibility
\( n=1 \) means \( \lambda =\mu ^{2}_{0} \) which is where the continuous spectrum
starts.)

\subsubsection{The continuous spectrum}

The continuous spectrum covers the range \( \mu ^{2}_{0}\leq \lambda <\infty  \).
In this domain we define a solution with the following asymptotic property:
\[
\Psi (\tau \to \infty )=e^{ip\tau }\,,\quad \lambda =\mu ^{2}_{0}+p^{2}\, .\]
Then one can compute 
\begin{equation}
\label{other_asymptotics}
\Psi (\tau \to -\infty )=\frac{\Gamma \left( \frac{ip}{\mu _{0}}\right) \Gamma \left( 1-\frac{ip}{\mu _{0}}\right) }{\Gamma (-s)\Gamma (1+s)}e^{-ip\tau }+\frac{\Gamma \left( -\frac{ip}{\mu _{0}}\right) \Gamma \left( 1-\frac{ip}{\mu _{0}}\right) }{\Gamma \left( -\frac{ip}{\mu _{0}}-s\right) \Gamma \left( 1+s-\frac{ip}{\mu _{0}}\right) }e^{ip\tau }\,.
\end{equation}
For \( s\in {\mathbb N} \) the first term vanishes, which means that the potential
is reflectionless. For \( s=1 \) we obtain
\begin{equation}
\label{phaseshift}
\Psi (\tau \to -\infty )=e^{i(p\tau +\delta (p))}\,,\quad e^{i\delta (p)}=\frac{ip+\mu _{0}}{ip-\mu _{0}}\,.
\end{equation}
To get a well-defined spectral density, we must put the system in a large box
of size \( T \). Then periodic boundary condition on \( \Psi  \) implies
\[
pT-\delta (p)=2\pi N\,,\quad N\in {\mathbb Z}.\]
 For the size \( T\to \infty  \) the density of states with a given \( p \)
is 
\begin{equation}
\label{density_definition}
\rho (p)=\frac{1}{2\pi }\left( T-\frac{\partial \delta (p)}{\partial
p}\right)\: , 
\end{equation}
which gives
\begin{equation}
\label{Q_density}
\rho (p)=\frac{1}{2\pi }\left( T-\frac{2\mu _{0}}{p^{2}+\mu ^{2}_{0}}\right) \: .
\end{equation}
A similar, but much simpler reasoning for \( Q_{0} \) yields
\begin{equation}
\label{Q0_density}
\rho _{0}(p)=\frac{T}{2\pi }\: ,
\end{equation}
so the divergent parts linear in \( T \) drop out from the difference.

\subsection{The instanton contribution}

Substituting the results (\ref{Q_density},\ref{Q0_density}) into (\ref{heat_kernel}),
the final result for the \( \zeta  \)-regularized heat kernel (\ref{zeta_regularization})
is
\[
\zeta (z,M)=\zeta _{1}(z,M)+\zeta _{2}(z,M)+\zeta _{3}(z,M)\,,
\]
where

\begin{eqnarray*}
\zeta _{1}(z,M) & = & \frac{1}{\Gamma (z)}\int ^{\infty }_{0}\ud t\, t^{z-1}L(4\pi t)^{-1/2}\left( K_{t}(Q)-K_{t}\left( Q_{0}\right) -1\right)\,, \\
\zeta _{2}(z,M) & = & \frac{1}{\Gamma (z)}\int ^{\infty }_{0}\ud t\, t^{z-1}\left( K_{t}\left( -\partial ^{2}_{x}\right) -1\right)\,, \\
\zeta _{3}(z,M) & = & \frac{1}{\Gamma (z)}\int ^{\infty }_{0}\ud t\, t^{z-1}\left( K_{t}\left( -\partial ^{2}_{x}\right) -L\left( 4\pi t\right) ^{-1/2}\right) \left( K_{t}(Q)-K_{t}\left( Q_{0}\right) -1\right)\,. 
\end{eqnarray*}
\( \zeta _{1} \) and \( \zeta _{3} \) correspond to renormalizing the classical
soliton mass \( M_{\mathrm{class}} \) to the quantum one. The interesting contribution
to the determinant comes from the term \( \zeta _{2} \) which gives
\begin{eqnarray*}
\left. \frac{\ud\zeta _{2}(z,M)}{dz}\right| _{z=0}&=&-\log
\frac{L^{2}}{4\pi }+\Gamma '(1)-2+\int ^{\infty }_{1}\frac{\ud
t}{t}\left( 1+\sqrt{\frac{4\pi t}{L^{2}}}\right) \left( K_{t}\left(
-\partial ^{2}_{x}\right) -1\right)\\
 &=&B-\log \frac{L^{2}}{4\pi }\: .
\end{eqnarray*}
Numerical evaluation of the integral gives
\[
2e^{\frac{B}{2}}=\frac{1}{\sqrt{\pi }}\]
to very high precision. Collecting all terms we get
\begin{equation}
\label{dilute_gas_result}
\Delta E(\vartheta )=-M\sqrt{\frac{2}{\pi l}}e^{-l}\cos \vartheta \, .
\end{equation}

\end{document}